\documentclass[fleqn,usenatbib]{mnras}

\usepackage{newtxtext,newtxmath}

\usepackage[T1]{fontenc}
\usepackage{ae,aecompl}

\usepackage{graphicx}	
\usepackage{amsmath}	

\usepackage{hyperref}   
\usepackage{gensymb}    
\usepackage{subcaption}
\usepackage{tabularx}
\usepackage{xcolor}
\usepackage{newtxtext,newtxmath}
\usepackage{lineno}

\defcitealias{Shanks_SMGs}{Paper I}

\title[An ALMA comparison of SMG dust heating mechanisms]{The nature of sub-millimetre galaxies II: an ALMA comparison of SMG dust heating mechanisms}

\author[B. Ansarinejad et al.]{
B. Ansarinejad$^{1,2}$\thanks{E-mail:behzad.ansarinejad@unimelb.edu.au (BA)},
T. Shanks$^1$\thanks{E-mail: tom.shanks@durham.ac.uk (TS)},
R.M. Bielby$^{1,3}$,
N. Metcalfe$^1$,
L. Infante$^{4,5,6}$,
D.N.A. Murphy$^7$
\newauthor
D.J. Rosario$^1$
\&
S.M. Stach$^1$
\\
$^1$Centre for Extragalactic Astronomy, Department of Physics, Durham University, South Road, Durham DH1 3LE, UK\\
$^2$School of Physics, University of Melbourne, Parkville, VIC 3010, Australia\\ 
$^3$Data Insights and Statistics Division, Department for Education, Bishopsgate House, Darlington, DL1 5QE, UK\\
$^4$Las Campanas Observatory, Carnegie Institution for Science, Colina El Pino S/N, La Serena, Chile\\
$^5$Nucleo de Astronomía de la Facultad de Ingeniería y Ciencias, Universidad Diego Portales, Av. Ejercito Libertador 441, Santiago, 8320000, Chile\\
$^6$Instituto de Astrofisica, Facultad de Fisica, Pontificia Universidad Catolica de Chile, Santiago, Chile\\
$^7$Institute of Astronomy, Madingley Road, Cambridge CB3 0HA, UK\\
}

\date{Accepted 2021 December 28. Received 2021 December 17; in original form 2021 July 1.}

\pubyear{2021}

\begin{document}
\label{firstpage}
\pagerange{\pageref{firstpage}--\pageref{lastpage}}
\maketitle

\begin{abstract}
{We compare the contribution of Active Galactic Nuclei (AGN)
and star-formation towards dust heating in sub-mm galaxies (SMGs).} We
have used ALMA at $0.''1$ resolution to image a {complete
flux-limited sample of {seven} sub-mm sources} previously shown 
to have spectral energy distributions (SEDs) that were as well-fitted by
obscured AGN as star-forming galaxy templates. Indeed, two sub-mm
sources were known to be quasars from their absorbed X-ray emission.
{We find the sub-mm sizes of all SMGs to be small}
($\approx1-2$kpc) and generally {$\sim3$ times} smaller than
any host detected in the {Near-Infra-Red (NIR)}. In all
cases, the {five} SMGs are comparable in sub-mm size to the
two known quasars and four $z\approx6$ quasars, also observed with ALMA.
We detect no evidence of diffuse spiral arms in this complete  sample.
We then convert the Far-Infra-Red (FIR) luminosities to star-formation
rate (SFR) surface densities and find that the SMGs occupy the same
range as the known quasars in our sample. We conclude that in terms of
sub-mm size, extent relative to host  and SFR density as well as
luminosity and {Mid-IR (MIR)} colour, there is little
distinction between the SMGs and  sub-mm bright quasars.
{Finally, we present preliminary evidence that SMGs with
higher MIR luminosities and sub-mm loud quasars tend to have dust
components that range to hotter temperatures  than their less luminous
SMG counterparts. In light of these results, we continue to suggest that
luminous SMGs may host dust-absorbed quasars that  may simultaneously
dominate the FIR and hard X-ray backgrounds.} \end{abstract}

\begin{keywords}
submillimetre: galaxies -- (galaxies:) quasars: supermassive black holes -- galaxies: starburst
\end{keywords}


\section{Introduction}

Sub-millimetre galaxies (SMGs) were first detected as highly luminous
{FIR} sources (see, e.g. \citealt{Smail1997};
\citealt{Barger1998}; \citealt{Hughes1998}; \citealt{Dey1999}) using the
James Clerk Maxwell Telescope's (JCMT) Submillimetre Common-User
Bolometer Array (SCUBA; \citealt{Holland1999}). These objects were soon
found to be high-redshift, dust-obscured sources which contribute to a
significant fraction {($\approx50$\%)} of the energy output of all galaxies in the early
Universe \citep{Blain2002}. Sub-millimetre observations, therefore,
opened a new window for studies of galaxy formation and evolution since
the cosmic dawn (see \citealt{Casey2014} for a detailed review).

{The identification of the dominant fuelling mechanism which
powers SMGs is an ongoing topic of research. The standard view is that
SMGs are predominantly luminous ($L_{\rm IR} > 10^{11} \textup{L}_\odot$) starburst
galaxies {(e.g. \citealt{Sanders1996})}, seen during an
obscured phase of their evolution (see, e.g.  {
\citealt{Sanders1996}}; \citealt{Alexander2005};
\citealt{Dudzeviciute2020}). However, it was also found that their high
apparent  star formation rates (SFR)  implied such high stellar masses
at such an early epoch, $z\approx2$, that they presented a problem for
the `bottom-up` $\Lambda-$Cold Dark Matter ($\Lambda$CDM)
model. Thus the Semi-Analytic Model of
\cite{Baugh2005} significantly under-predicted the abundance of SMGs,
requiring the somewhat ad-hoc adoption of a  top-heavy stellar initial
mass function in starbursts to increase their luminosity compared to
their mass and thus account for the high observed SMG number counts
without contradicting the $\Lambda$CDM mass function (see also e.g.
\citealt{Geach2017}; \citealt{Cowley2019}).}

An alternative view is that Active Galactic Nuclei (AGN) may be the
dominant mechanism that powers luminous SMGs (e.g. \citealt{Hill2011a}).
Although 10-20\% of SMGs are accepted to host AGN (\citealt{Cowie2018,Franco2018,Stach2019}), 
they are usually viewed to be sub-dominant to star-formation in heating these sources
(see, e.g. \citealt{Laird2010}; \citealt{Johnson2013};
\citealt{Wang2013}).

On the other hand, there are various arguments for considering obscured
AGN as the primary power source for {at least, bright, 
$S_{870\rm \micro m}\ga 1$ mJy, SMGs.} For instance, a population of
heavily obscured quasars could explain the missing hard X-ray background
(\citealt{Comastri1995}; \citealt{Worsley2005}; \citealt{Polletta2007}),
while their dust-rich nature would mean that they would have large
emissions in the Infra-Red (IR) due to the emission of reprocessed light
from the AGN. Indeed, obscured AGN models have been shown to provide a
reasonable fit to the bright end of the SMG source counts
\citep{Hill2011a}, reducing the need for a top-heavy IMF for
high-redshift starbursts, although issues about the origin of the dust
in the AGN may still  remain (C.G. Lacey, priv. comm.). Note also that,
at fainter flux densities, star-forming galaxies are still expected to dominate
the observed SMG number counts. Finally, for AGN to be the dominant
source of powering the sub-mm emission rather than star formation, the
dust torus must lie far enough (at $\approx$ kiloparsec scales) from the
nucleus to maintain a cool temperature of $\approx35$K and produce
spectra consistent with observations; a picture which is feasible if one
assumes a torus model similar to e.g. \cite{Kuraszkiewicz2003}.   

In recent years, the unprecedented sensitivity and angular resolution of
Atacama Large Millimetre/Submillimetre Telescope (ALMA) has enabled the
study of the dust heating mechanisms of SMGs at significant redshifts.
Using $0.''3$ imaging from ALMA, \cite{Simpson2015} found that most of
their targeted $z\approx2$ SMGs are just resolved, with their imaging
probing scales of $\approx 2-3$ kpc. Furthermore, they found the
$K$-band optical extent of these SMGs to be roughly four times larger
than their extent in sub-mm. Using higher resolution ($0.''16$) ALMA
imaging, \cite{Hodge2016} found sub-mm sizes of $\approx1.3$ kpc for 16,
$S_{870 \rm \micro m}\approx 3-9$ mJy, SMGs, with a selection skewed toward the
most luminous of the 122 sources in the `ALMA follow-up of the LABOCA
ECDFS sub-mm survey' (ALESS). Similarly, \cite{Gullberg2019} studied a {stacked}
sample of $\sim150$ SMGs with $0.''18$ ALMA resolution, finding a
compact sub-mm dust continuum emission extended to just $\sim1$kpc, in
comparison to the Hubble Space Telescope (HST) imaging of the optical/UV
emissions of the same sources extending to $\sim8-10$ kpc. {These results
may give some rough impression that the better the resolution, the
smaller the sub-mm extent that is measured. All studies appear to agree
that the sub-mm extent is significantly smaller than the galaxy host in
the rest optical, although not all have the benefit of HST imaging. }

{Then higher resolution ($0.''08$) and, perhaps more
importantly, higher S/N ALMA observations by \citet{Hodge2019} found
evidence for small-scale (i.e 1-2 kpc radius) spiral arms in a randomly
selected (with respect to morphology) sub-sample of 6 of the above 16
luminous ALESS SMGs. More precisely, these 6 SMGs were selected as the
sub-mm-brightest sources from the 16 ALESS SMGs with previous
high-resolution ($0.''16$) 870 $\micro$m ALMA imaging from
\cite{Hodge2016}, which were themselves chosen as the sub-mm-brightest
sources with (randomly targeted) HST coverage.} These authors attributed
the compact size of the sub-mm emitting region of these sources to a
central starburst evolving into a galactic bulge or bar. However, an
alternative interpretation could be the presence of AGN which are
heating the inner regions of these SMGs, producing luminosities
comparable to quasar emissions and it is this AGN hypothesis we aim to
test further here.

In this paper, we analyse a complete flux-limited sample of 7 sub-mm
sources including 5 unidentified $z\approx2$ SMGs and  2 X-ray absorbed
quasars located in the William Herschel Deep Field (WHDF, e.g.
\citealt{Metcalfe2006}). These sub-mm sources were originally detected
by APEX LABOCA \citep{Bielby2012} and then targetted by ALMA. We also
include a sixth fainter SMG, detected by  ALMA near one of the quasars.
We further include 4 $z>6$ quasars originally identified in the VST
ATLAS survey \citep{Shanks2015,Carnall2015,Chehade2018} that ALMA has
also detected as sub-mm sources.

{The WHDF sample is well-suited for comparing AGN and star-forming heat sources
since it has the combination of high resolution to measure the size of nuclear 
features and long 1500s exposures to maximise the chance of detecting low surface brightness features
such as spiral arms for each target. It also includes two sources that are already known to be QSOs 
from their X-ray emission and optical spectra that can act as AGN templates. The sample 
of 5 sources is also complete in the central WHDF area to a fixed flux density limit.
The 4 $z>6$ quasars act as further high luminosity QSO templates with similar kpc scale resolution 
and similarly long ALMA exposures.}

In the case of the  WHDF sub-mm sources, \citet{Shanks_SMGs}
(hereafter \citetalias{Shanks_SMGs}) have already compared AGN and
star-forming fits to the spectral energy distributions (SEDs) of these
sources, constructed using multi-wavelength data ranging from X-ray to
radio bands. There, it was found that AGN SEDs fitted most SMGs as well
as star-forming galaxy templates and the SMG MIR colours were generally
indistinguishable from quasars. Here, we present further comparisons
between the SMGs and the quasars of their FIR luminosities, sizes and 
host galaxy relative extents based on our ALMA FIR continuum imaging.
The high ALMA resolution and S/N is again competitive with the best previous
studies  and is  clearly advantageous for measuring these properties and
making these comparisons.  

Measuring the star formation rate surface density,
$\sum_{\textup{SFR}}$, of SMGs is an additional method of comparing our
SMG and quasar sub-samples. $\sum_{\textup{SFR}}$ can also test whether
the observed SFR surface density anywhere exceeds the `Eddington limit'
of star formation set by radiation pressure on dust\footnote{The
Eddington limit on $\sum_{\textup{SFR}}$ arises due to the momentum
deposited due to radiation pressure from stars blowing the star-forming
gas and dust out of the system. The value quoted here is a lower bound
as estimated by \cite{Hodge2019}.} ($\sum_{\textup{SFR}}\approx650~{\rm
M}_\odot~{\rm yr}^{-1}~{\rm kpc}^{-2}$; see, e.g.
\citealt{Thompson2005}; \citealt{Walter2009}; \citealt{Hopkins2010};
\citealt{Decarli2018}; \citealt{Hodge2019}), perhaps  ultimately
requiring an alternative heating mechanism such as AGN.

The outline of this paper is as follows; in Section~\ref{sec:SMG_obs} we
describe the observations of our WHDF SMGs, X-ray quasars and $z>6$
quasars. In Section~\ref{sec:SMG_results}, we describe our measurements
of the sizes and fluxes of our ALMA sources as well as their host
galaxies. In Section~\ref{sec:SMG_SFRs} we derive FIR luminosities and
estimate SFR surface densities based on continuum fluxes and extents. In
Section~\ref{sec:SMG_discussion} we summarise our comparison of the FIR
properties of SMGs and quasars before presenting our conclusions in
Section~\ref{sec:SMG_conclusions}. Throughout, we adopt a fiducial 
cosmology with $H_0=70$ km s$^{-1}$ Mpc$^{-1}$,
$\Omega_\textup{m}=0.3$, $\Omega_\Lambda=0.7$. {Finally, note that for calculating SFR's we
assume a Chabrier IMF.}

\section{Observations}
\label{sec:SMG_obs}

\subsection{Initial FIR detection of  WHDF SMGs and quasars}
\label{sec:whdf_obs}

Initially, 11 sub-mm sources were  detected in an APEX LABOCA
\citep{Siringo2009} $S_{870\textnormal{\micro m}}>3.3$ mJy
survey \citep{Bielby2012}. The {seven sources subsequently
targeted by ALMA, together with an eighth, LAB-06, form a complete
flux density limited sample located in the central $7'\times7'$ area of the
WHDF. LAB-06 was not targeted by ALMA because it is was already
identified with a low redshift ($z=0.046$) spiral galaxy.} Two of these
sources LAB-05 and LAB-11 have been identified as X-ray absorbed quasars
by \cite{Bielby2012} with redshifts of $z=2.12$ and $z=1.32$
respectively. A summary of the redshifts and coordinates of these
sources can be found in Table~\ref{tab:QSO_summary}. The redshifts for
LAB-05 and LAB-11 are spectroscopic, the other 5 are from the AGN SED
fits of \citetalias{Shanks_SMGs}; these have considerably larger errors,
{typically $\approx12$\%.} However, the well known dependence
of angular diameter distance on redshift means that any redshift with
$0.4<z<6.3$ must have a scale within the range 0.54-0.85 kpc / $0.''1$.
So for a $0.''1$ angular scale, the physical scale peaks at
$z\approx1.6$ at 0.85 kpc while a galaxy at $z=6.3$ is as well resolved
as a galaxy at $z=0.4$ with a physical scale of 0.54 kpc.

\subsection{Initial detection of $z>6$ ATLAS quasars}
\label{sec:z6_obs}

The four $z>6$ quasars studied here were initially identified as quasar
candidates in the VST ATLAS survey \citep{Shanks2015} using the colour
selection criteria of \cite{Carnall2015} and \cite{Chehade2018}. These
candidates were confirmed as $z\approx6$ quasars using low-resolution
spectroscopy on the LDSS-3 instrument on the Magellan 6.5m telescope,
and then with moderate resolution X-Shooter \citep{Vernet2011} spectra
by \cite{Chehade2018}. A summary of the properties of these quasars
including their black hole masses inferred from the [C{\sc iv}] broad
emission-line width $M_{\textup{BH}}$[$\textup{C}_{\textup{IV}}$]
reported by \cite{Chehade2018} is presented in
Table~\ref{tab:QSO_summary}. These luminous quasars will be of interest
here mainly for comparison with the WHDF SMGs and lower redshift
quasars.

\subsection{ALMA observations of the WHDF LABOCA sources}
\label{sec:SMG_ALMA_obs}

{The  flux-limited subset of seven} LABOCA sources  in the central
$7'\times7'$ of the WHDF were  targeted  with   ALMA  in Band 7
(275–373GHz) on 11/10/2016 with the 12-m Array   in a configuration
which yielded 870 $\micro$m continuum images at $0.''095$ resolution and a
maximum recovered scale of $0.''926$. {The median precipitable water
vapour (PWV) at zenith was recorded as 0.65mm.} LAB-01, -02,
-03, -04, -05, -06, -10, -11 formed the complete {flux-limited} sample but we excluded
LAB-06 as an ALMA target on the grounds that it was already identified
with a nearby, $z=0.046$, spiral galaxy. The other {seven} sources thus
comprise our flux-limited sample observed by ALMA. The exposure
times were 1572s each, aimed at being long enough to detect any diffuse
emission (e.g. spiral arms) surrounding the sub-mm core. These
observations reached an 870 $\micro$m surface brightness rms {
of 60 $\micro$Jy per $0.''09\times0.''11$ beam over a $\approx17''$ diameter
field-of-view. This can be compared to diffuse  spiral arms of
\cite{Hodge2019} having a surface brightness of typically 200$\micro$Jy per
$0.''08\times0.''06$ beam or 400$\micro$Jy/beam at our resolution. Thus
similar spiral features should be detected by our observations at
$\approx7\sigma$/beam, less than the $\approx9\sigma$/beam of
\cite{Hodge2019} but at high significance nonetheless.}

 
To calibrate the observational data, we make use of the standard ALMA data reduction and calibration scripts provided with the raw observations. These scripts were ran with the {\sc Common Astronomy Software Application} ({\sc CASA} v4.7.2; \citealt{McMullin2007}) package and we do not perform any additional tapering or cleaning\footnote{Note that the ALESS images that we show in Fig.
\ref{fig:lab_sources}(c) have the same pipeline reduction consistently
applied as for the WHDF images in Fig. \ref{fig:lab_sources}(a).}. All seven sources were strongly
detected with LAB-11 revealing a companion at $\approx5''$ from the main
LAB-11 source {that had been unresolved in the LABOCA data},
making an eighth ALMA sub-mm source in the central WHDF area, now named
LAB-12. {Although this source has an 870 $\micro$m flux density below
3.3mJy as indeed does LAB-11 we shall continue to include these  in our
main sample, since for our purposes it is more important that our sample
is demonstrably unbiased in terms of morphology rather than being specifically flux
density limited.}

\begin{figure}
   \centering
     \includegraphics[width=\columnwidth]{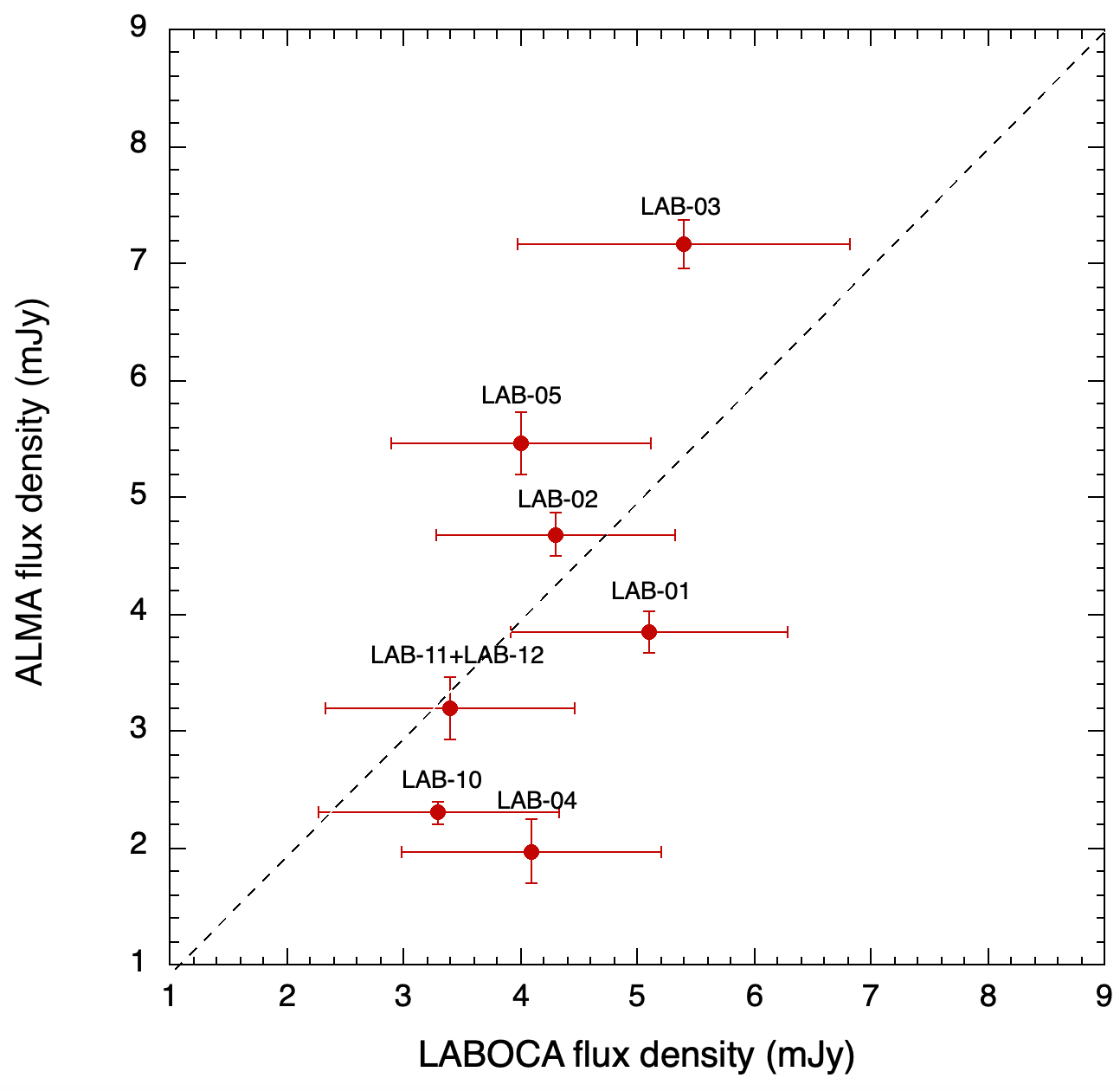}
	\caption{Comparison of ALMA and LABOCA $870 \rm \micro m$ flux densities for the WHDF SMGs.
	{LAB-11 + LAB-12 flux densities for ALMA are summed in this comparison as they were unresolved} in
	the LABOCA imaging of \protect\cite{Bielby2012}.}
	\label{fig:ALMA_v_lab}
\end{figure}

\begin{figure*}
    \centering
	\begin{subfigure}{\textwidth}
	\centering
	    \includegraphics[width=0.9\textwidth]{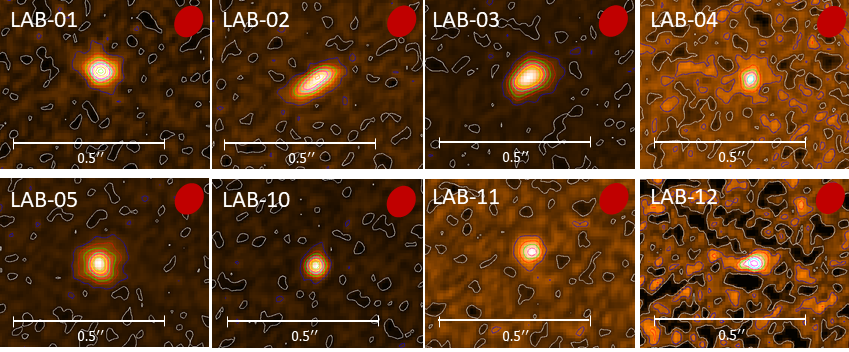}
	    \caption{}
		\label{fig:lab_sources_ALMA}
	\end{subfigure}
	\centering
	\centering
	\begin{subfigure}{\textwidth}
	\centering
	    \includegraphics[width=0.45\textwidth]{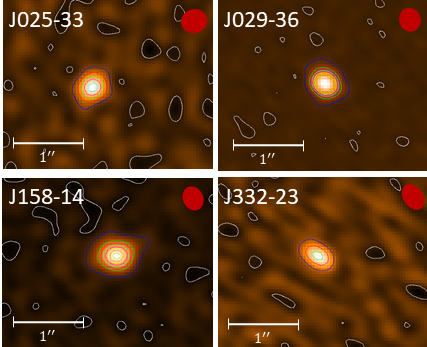}
		\caption{}
		\label{fig:hiz_sources_ALMA}
	\end{subfigure}
	\begin{subfigure}{\textwidth}
	\centering
	    \includegraphics[width=0.7\textwidth]{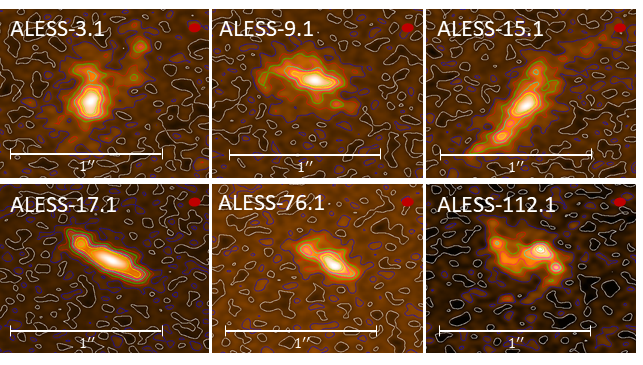}
		\caption{}
		\label{fig:ALESS_sources_ALMA}
	\end{subfigure}

	\caption[ALMA images and profile fits for our $z\approx2$ SMGs]{(a)
	ALMA `thumbnails' with a resolution of $0.''1$, showing the
	continuum emitting regions of the WHDF sub-mm sources. (b)
	ALMA `thumbnails' with a resolution of $0.''3$, showing the
	continuum emitting regions of the $z\approx6$ ATLAS quasars. (c) $0.''07$ resolution
	ALMA `thumbnails showing the continuum emission and hints of
	sub-structure detection in the ALESS SMGs of \cite{Hodge2019}. In all cases,
	the contour levels are kept consistent to allow for direct visual
	comparison of the extent of the sources {and the beams are shown by the red ellipses}.} 
	\label{fig:lab_sources}
\end{figure*}

In order to ensure our ALMA observations have recovered the full flux
(and therefore the full extent) of the WHDF sources, in
Fig.~\ref{fig:ALMA_v_lab} we compare the ALMA flux density of these objects with
previous LABOCA flux density measurements of \cite{Bielby2012}. {
Here we have treated LAB-11 and LAB-12 as a single object, summing their
ALMA flux densities to find their total ALMA flux density of
$3.19\pm0.27$ mJy before plotting against the LABOCA flux density for
LAB-11 of $3.4\pm1.06$ mJy. With the possible} exception of LAB-04 where
the LABOCA flux density appears to be higher than the ALMA flux density (albeit at a 
$<2\sigma$ level of significance), the measurements for the remaining
sources are in agreement within the error on the LABOCA flux density 
measurements. {In the case of LAB-04, we checked whether any low surface
brightness component was missed by trying a range of data smoothing
but such tests for LAB-04 and other targets revealed no such components.}
The lower ALMA flux could be due to the source being placed closer to
the edge of the observed ALMA field of view  where the ALMA sensitivity
and S/N is reduced.

\subsection{ALMA observations of the $z>6$ ATLAS quasars}
\label{sec:z6_ALMA_obs}
The four $z>6$ quasars were  targeted  with   ALMA  in Band 6
(211–275GHz) on 14, 17, 19 and 24/11/2016 with the 12-m Array in
configurations which yielded $\approx1150 \micro$m continuum images at
$\approx0.''35-0.''4$ resolution and a maximum recovered scale of
$3.''4-4.''0$. The exposure times for J029-36, J332-32, J025-33
and J158-14 {(SIMBAD IDs: VST-ATLAS J015957.96-363356.8; QSO J2211-3206; QSO J0142-3327; PSO J158.6937-14.4210)} were 2956, 1845, 1875 and 1996 seconds, reaching continuum
sensitivities of 0.020, 0.017, 0.020 and 0.014 mJy/beam. Although the
spatial resolution of these  ALMA observations is $\approx3-4\times$
lower than for the WHDF sources, we are able to probe proper distances
as small as those at $z\approx0.5$, due to the reduction of the angular
diameter distance, $d_\textup{A}(z)$, at redshifts $z>1.6$. Taking advantage of this
property of the Universe, we were able to resolve 3/4 of our $z\approx6$
SMGs in the dust continuum.



All four of our $z\approx6$ quasars were targeted by ALMA for [C{\sc
ii}] emission but small uncertainties in the `discovery' redshifts of two of
these sources (J332-23 and J029-36) resulted in the [C{\sc ii}] line
being missed. In the case of J025-33 the [C{\sc ii}] line was located at
the edge of our detection window missing some of the [C{\sc ii}] flux.
Although \cite{Decarli2018} have presented ALMA [C{\sc ii}]
observations of J025-33 (albeit with a lower angular resolution of
$0.''8$), in light of these issues we postpone discussion 
of the line measurements to future work. 

\begin{table*}
	\centering
	\caption[Properties of the SMGs targeted in our ALMA observations]{Summary of WHDF SMG and $z>6$ QSO properties.}
	\label{tab:QSO_summary}
	\begin{tabular}{lccccccc} 
		\hline
		Name & Short Name & RA & Dec & $z$  & $M_{1450\textup{\AA}}$ & $M_{\textup{BH}}$[$\textup{C}_{\textup{IV}}$] & Reference(s) \\
             &      & (J$2000$)&     &      & (Mag)                                                                  & ($10^9$\(\textup{M}_\odot\)) & \\
        (1)  &  (2) &      (3) & (4)      &(5)& (6)                  & (7)                                           & (8)\\
		\hline
        WHDF-LAB-01 & LAB-01 & 00:22:37.58 & +00:19:18.4  & $2.60\pm0.15$ & --- & --- & 3,4 \\[0.1cm]
        WHDF-LAB-02 & LAB-02 & 00:22:28.44 & +00:21:47.6  & $3.10\pm0.25$ & --- & --- & 3,4 \\[0.1cm]
        WHDF-LAB-03 & LAB-03 & 00:22:45.96 & +00:18:41.2  & $2.70\pm0.35$\dag & --- & --- & 3,4 \\[0.1cm]
        WHDF-LAB-04 & LAB-04 & 00:22:29.19 & +00:20:24.8  & $3.00\pm0.60$ & --- & --- & 3,4 \\[0.1cm]
        WHDF-LAB-05 & LAB-05 & 00:22:22.87 & +00:20:13.5  & $2.12\pm0.03$ & --- & --- & 3,4 \\[0.1cm]       
        WHDF-LAB-10 & LAB-10 & 00:22:35.23 & +00:24:07.5  & $0.90\pm0.20$\dag & --- & --- & 3,4 \\[0.1cm]
        WHDF-LAB-11 & LAB-11 & 00:22:24.84 & +00:20:11.4 & $1.32\pm0.03$ & --- & --- & 3,4 \\[0.1cm]
        WHDF-LAB-12 & LAB-12 & 00:22:25.48 & +00:20:06.6 & $2.90\pm0.10$ & --- & --- & 3,4 \\[0.1cm]
		\hline
		VST-ATLAS J332.8017-32.1036 & J332-23 & 22:11:12.41 & -32:06:12.96 & $6.32\pm0.03$ & $-26.79\pm0.06$ & 2.7 & 1 \\[0.1cm]
		VST-ATLAS J158.6938-14.4211 & J158-14 & 10:34:46.51 & -14:25:15.96 & $6.07\pm0.03$  & $-27.23\pm0.08$ & 2.4 & 1 \\[0.1cm]
		VST-ATLAS J025.6821-33.4627 & J025-33 &  01:42:43.70 & -33:27:45.72 & $6.31\pm0.03$ & $-27.50\pm0.06$ & 2.2 & 2 \\[0.1cm]
        VST-ATLAS J029.9915-36.5658 & J029-36 & 01:59:57.96 & -36:33:56.88  & $6.02\pm0.03$ & $-26.97\pm0.08$ & 1.4 & 2 \\[0.1cm]
        \hline
        \multicolumn{8}{p{\textwidth}}{Notes. (1) {Source name (these names match the publications were the source names were first introduced; see references in column 8)}, (2) Short
        Name, (3) Right Ascension, (4) Declination, (5) Inferred
        Ly$\alpha$ redshift in case of ATLAS $z\approx6$ quasars, and redshift
        based on AGN template SED fits presented in Table 2 of
        \protect\citetalias{Shanks_SMGs}. $\dag$: redshifts estimated based on SED
        fits using optical/MIR detections of close companions to LAB-03
        and LAB-10 since the direct counterpart of these sources are
        undetected in these bands. (6) $1450\textup{\AA}$ rest-frame
        absolute magnitude, (7) {Black hole} mass estimated from the C{\sc iv}
        broad emission-line width, (8) References:
         1.-\protect\cite{Chehade2018};
         2.-\protect\cite{Carnall2015}, 3.-\protect\cite{Bielby2012}, 4.-\protect\citetalias{Shanks_SMGs}
        .}
	\end{tabular}
\end{table*}

\section{SMG and quasar FIR source sizes and morphologies}
\label{sec:SMG_results}

\subsection{WHDF source sizes and morphologies via CASA IMFIT}
\label{sec:imfit}

In Figs.~\ref{fig:lab_sources_ALMA} and \ref{fig:hiz_sources_ALMA} we
show the FIR continuum emission detected in our ALMA observations of the
WHDF SMGs and ATLAS $z>6$ quasars respectively. We measure the apparent
and deconvolved sizes of the WHDF SMGs and ATLAS quasars {by
fitting Gaussians} using the {\sc
imfit}{\footnote{\url{https://casa.nrao.edu/docs/taskref/
imfit-task.html}}} routine, which is part of the {\sc CASA
(v4.7.2)} package. As well as a measurement of the extent of
the sources, {\sc imfit} provides the integrated and peak flux (per
beam) for each object which are used to calculate their SFR as detailed
in Section~\ref{sec:SMG_SFRs}. The measured sizes of the WHDF SMGs are
given in Table \ref{tab:lab_sizes}. We see that all are resolved at our
$0.''1$ resolution. 

\begin{figure*}
	\begin{subfigure}[H]{0.495\textwidth}
		\centering
		\includegraphics[width=\textwidth]{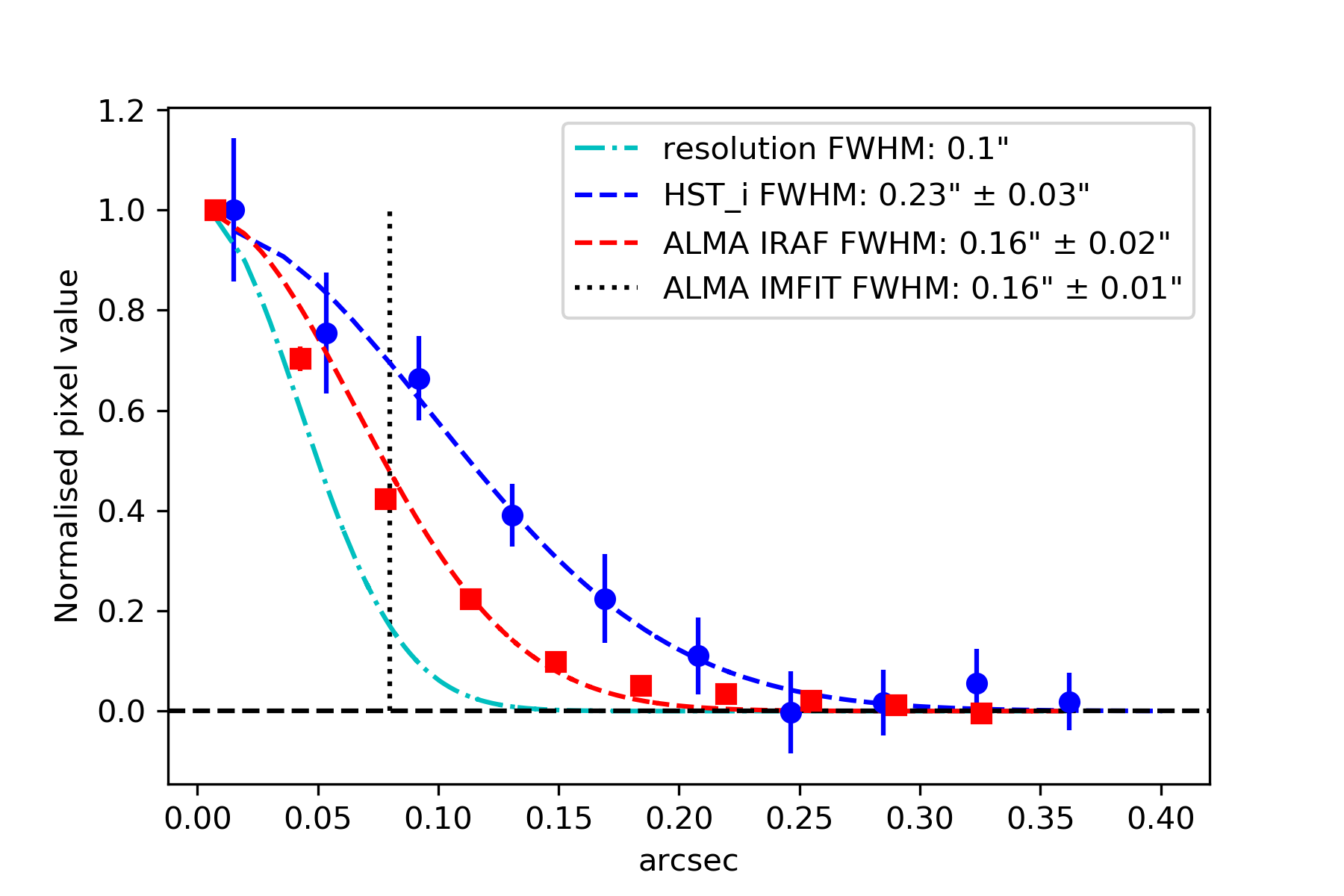}
		\caption{LAB-01}
		\label{fig:Lab1_i_hst}
	\end{subfigure}	
	\begin{subfigure}[H]{0.495\textwidth}
		\centering
		\includegraphics[width=\textwidth]{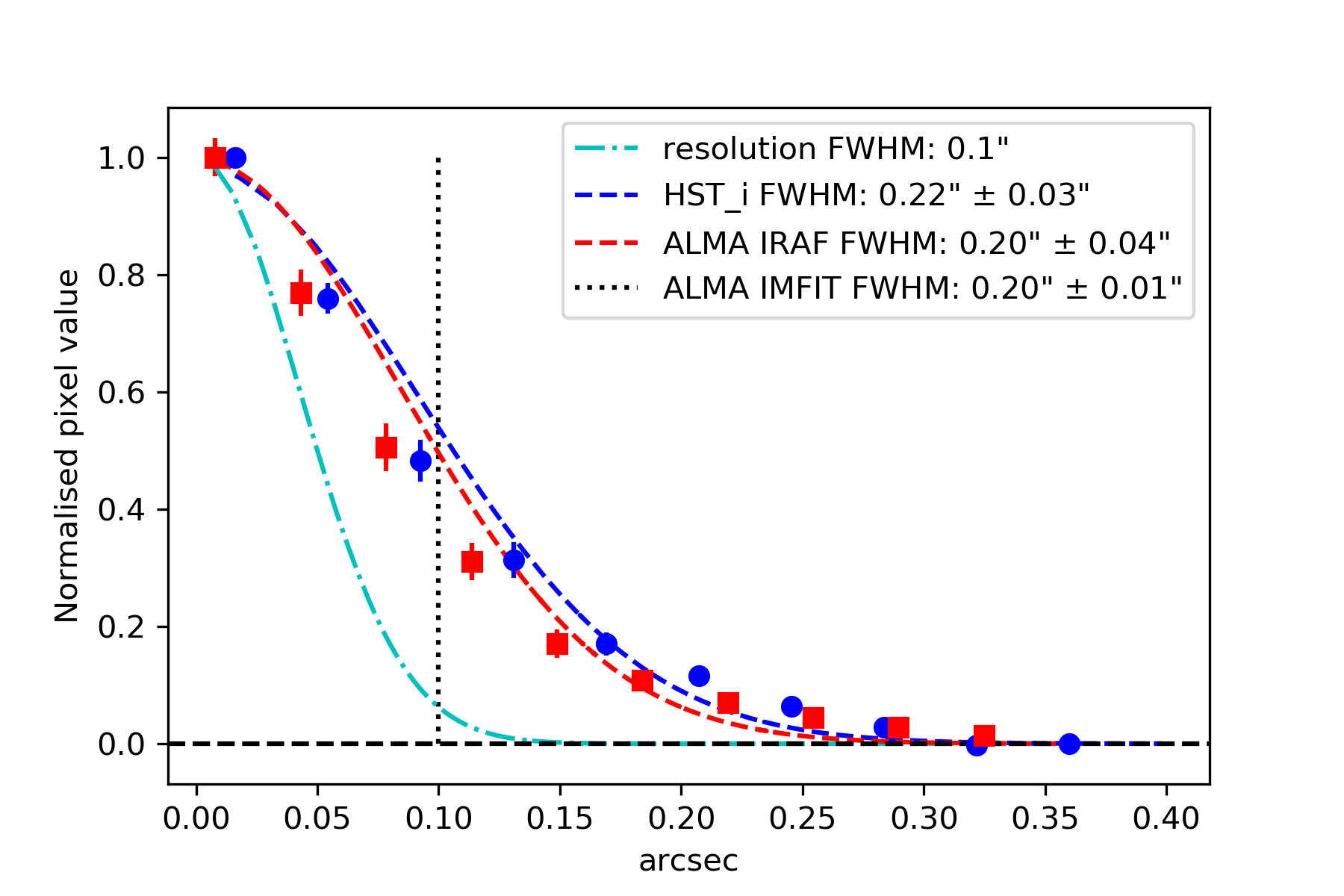}
		\caption{LAB-02}
		\label{fig:Lab2_i_hst}
	\end{subfigure}
	\begin{subfigure}[H]{0.495\textwidth}
		\centering
		\includegraphics[width=\textwidth]{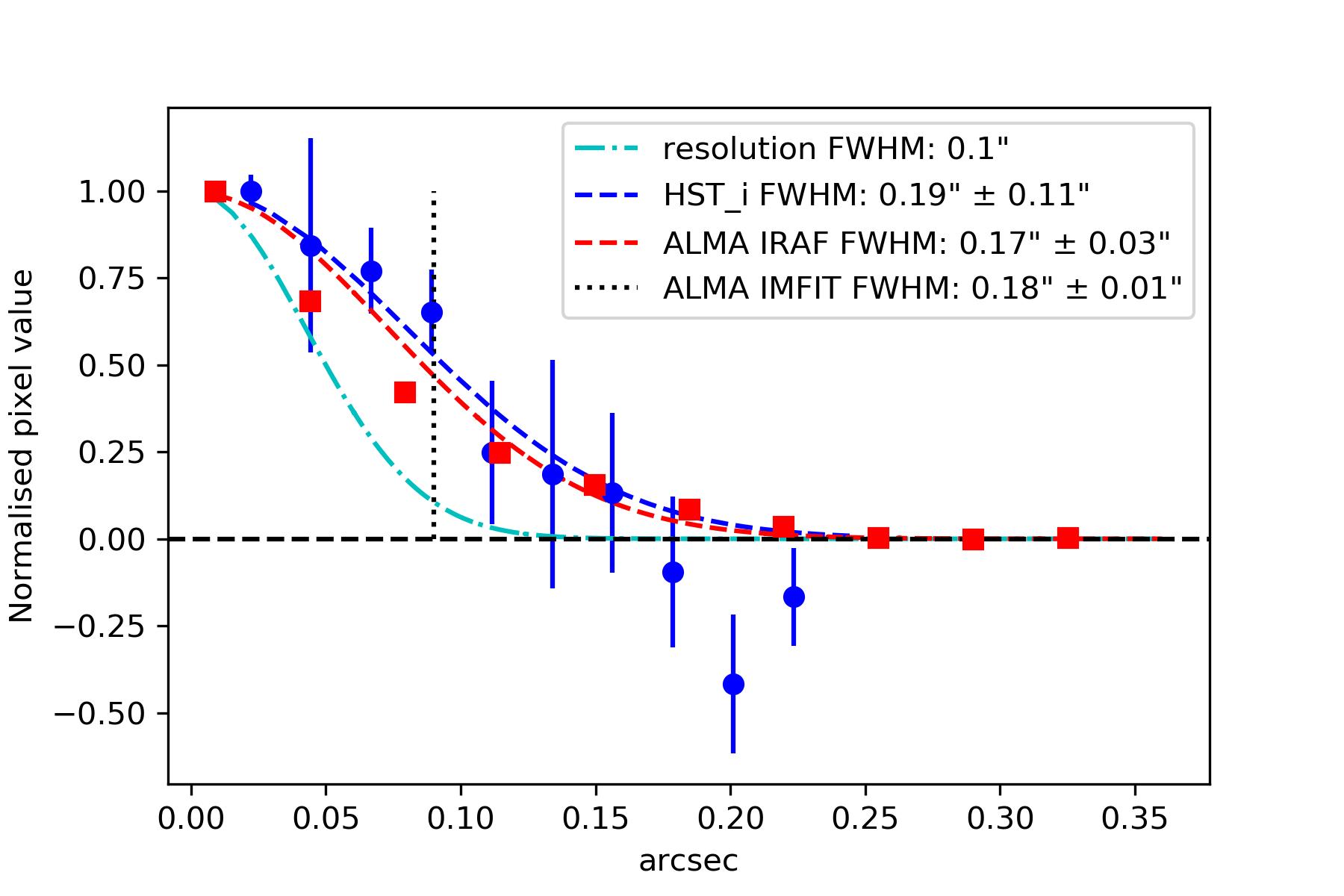}
		\caption{LAB-05}
		\label{fig:Lab5_i_hst}
	\end{subfigure}
	\begin{subfigure}[H]{0.495\textwidth}
		\centering
		\includegraphics[width=\textwidth]{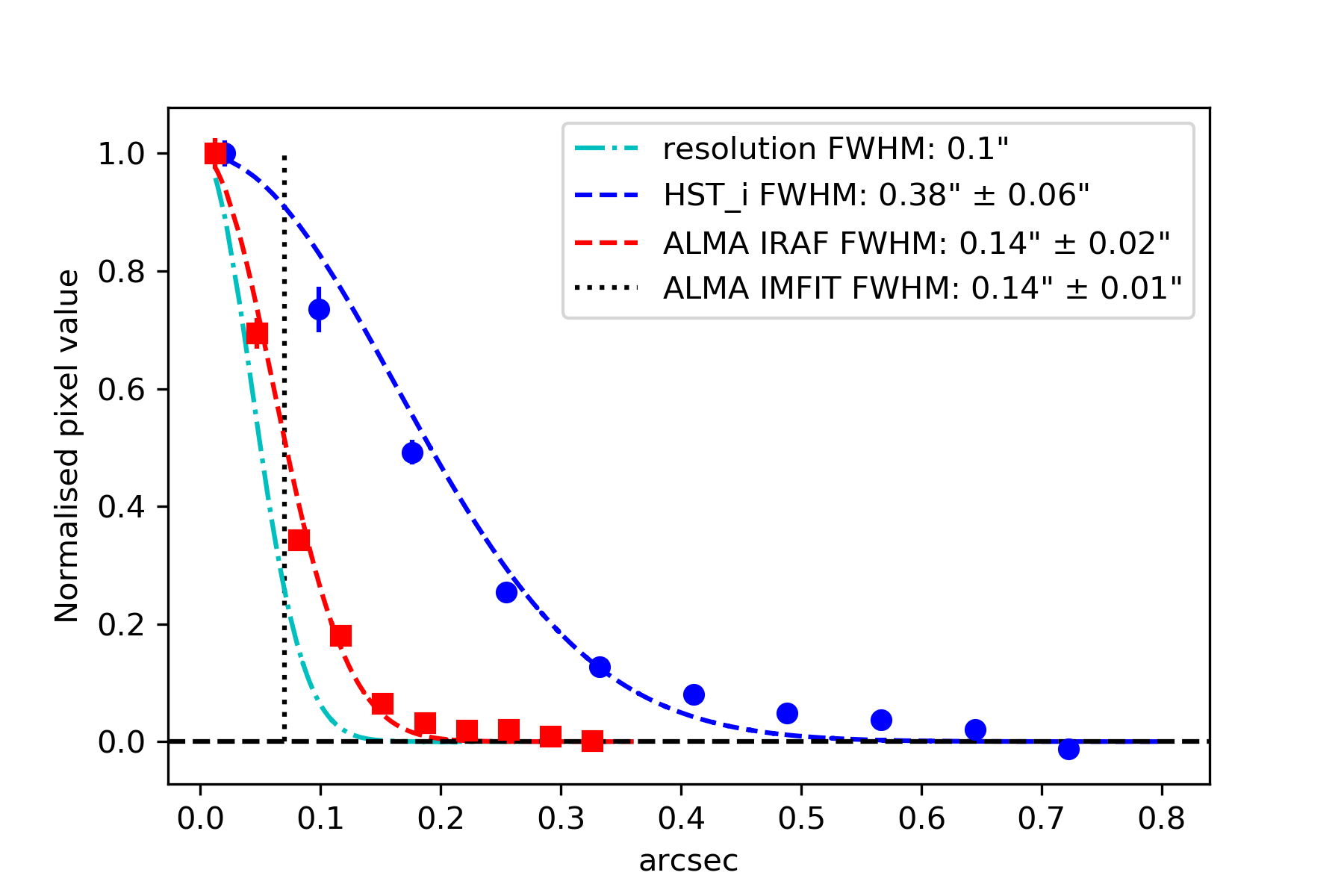}
		\caption{LAB-11}
		\label{fig:Lab11_i_hst}
	\end{subfigure}
	\begin{subfigure}[H]{0.495\textwidth}
		\centering
		\includegraphics[width=\textwidth]{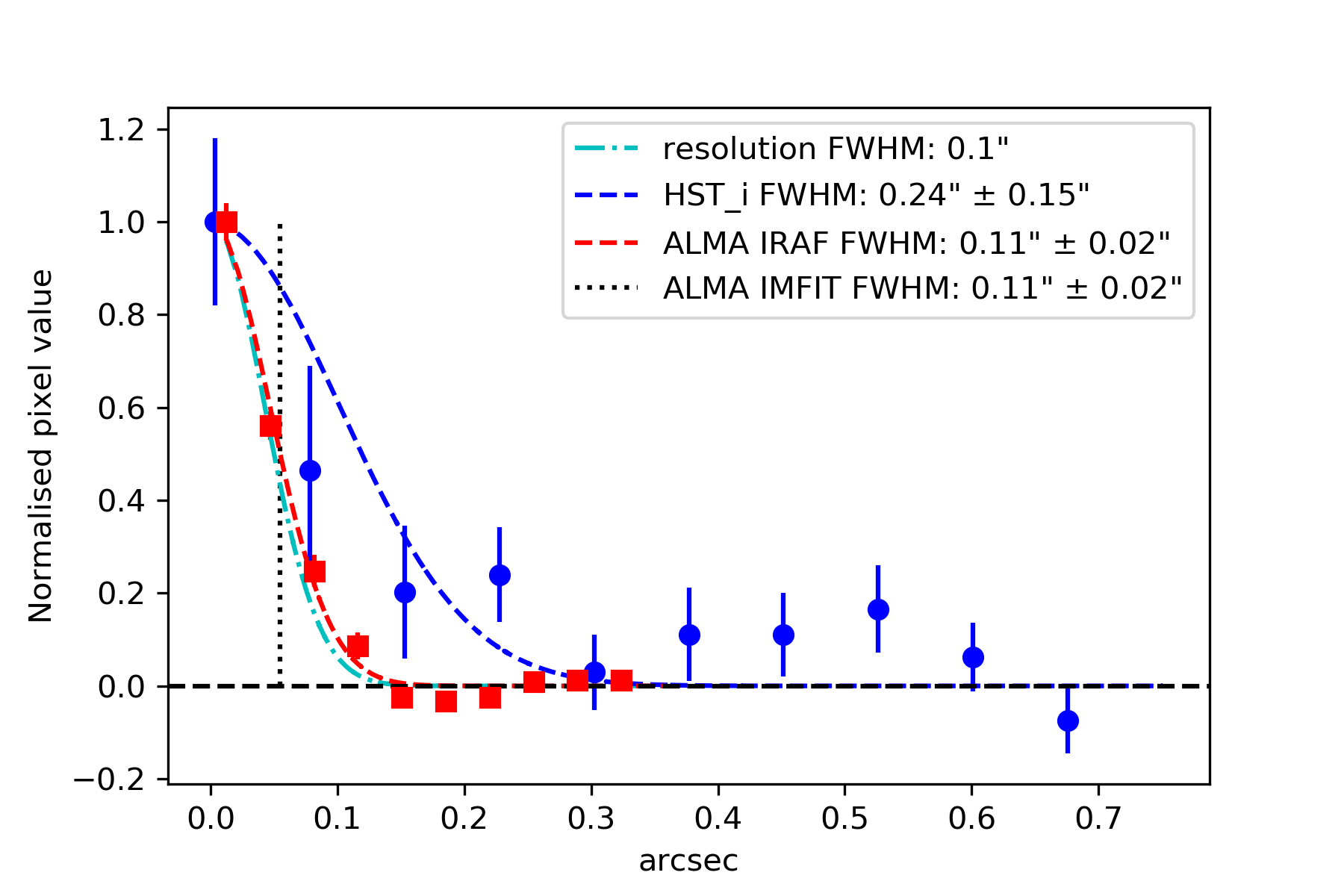}
		\caption{LAB-12}
		\label{fig:Lab12_i_hst}
	\end{subfigure}
	\caption{Comparison of ALMA and HST $i$-band profiles of LABOCA sources with detections in both bands. For comparison we include a Gaussian with a $0.''1$ FWHM indicating the resolution of the HST and ALMA imaging. Here, no PSF correction has been applied to the ALMA or HST profiles. The range of redshifts covered here is $1.32<z<3.1$ so the observed $i$-band corresponds to the rest wavelength range 2100-3700\AA~ which seems acceptably narrow. Indeed, excluding LAB-11 this range reduces still further to 2700-3700\AA, arguing that these $i-$band profiles can be consistently compared. }
	\label{fig:HST_i}
\end{figure*}

\begin{figure*}
	\begin{subfigure}{0.495\textwidth}
		\centering
		\includegraphics[width=\textwidth]{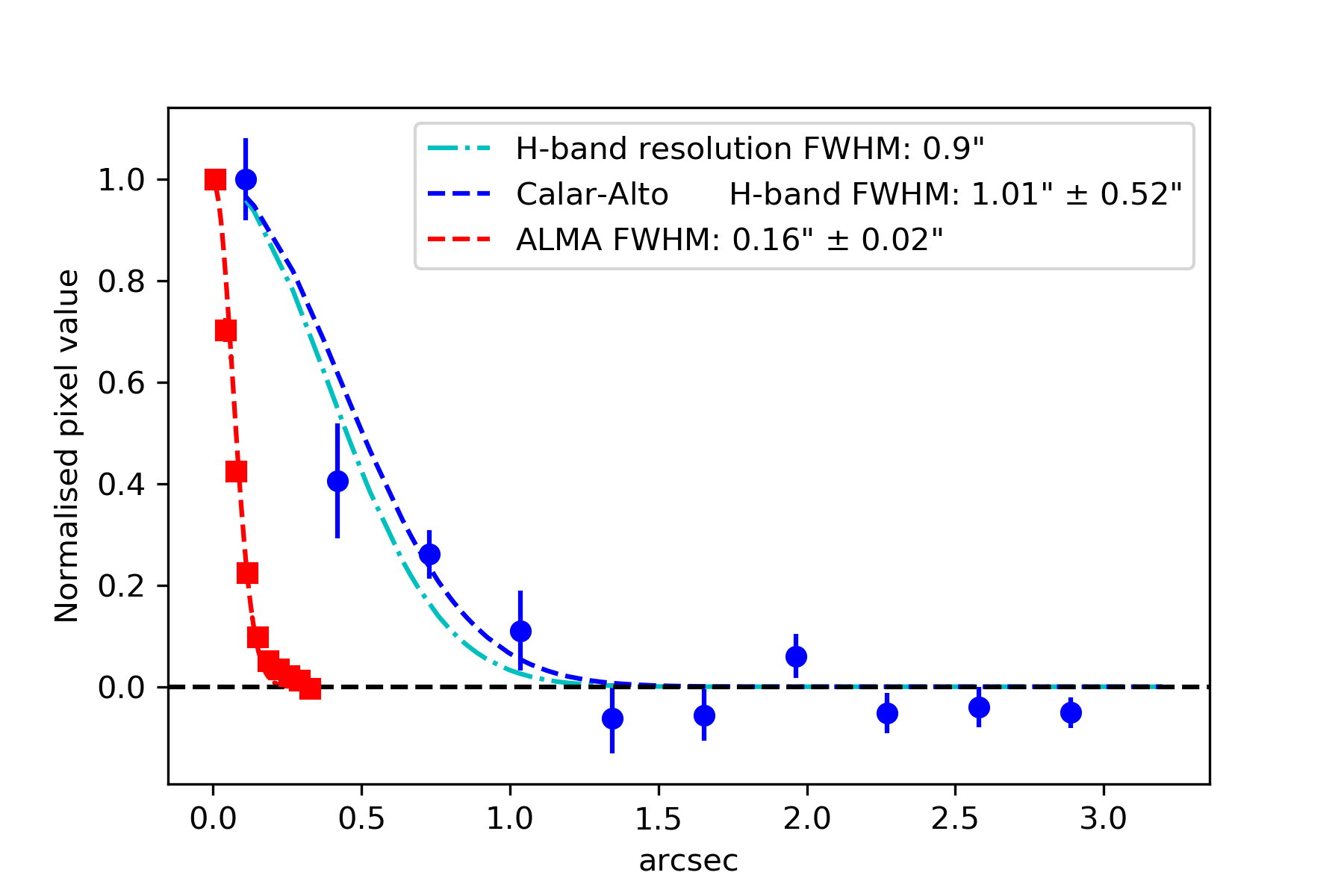}
		\caption{LAB-01}
		\label{fig:Lab1_h}
	\end{subfigure}	
	\begin{subfigure}{0.495\textwidth}
		\centering
		\includegraphics[width=\textwidth]{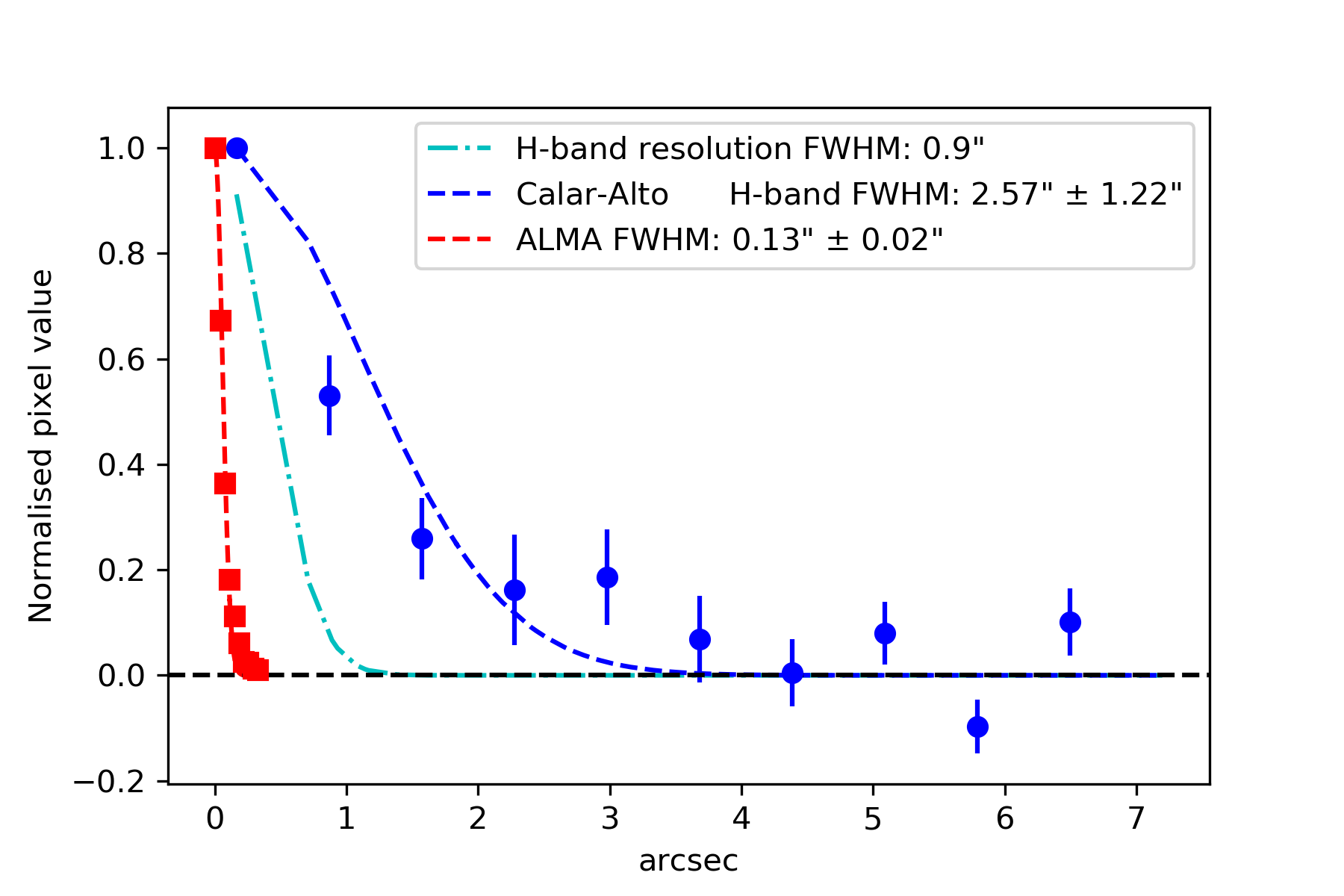}
		\caption{LAB-04}
		\label{fig:Lab4_h}
	\end{subfigure}
	\begin{subfigure}{0.495\textwidth}
		\centering
		\includegraphics[width=\textwidth]{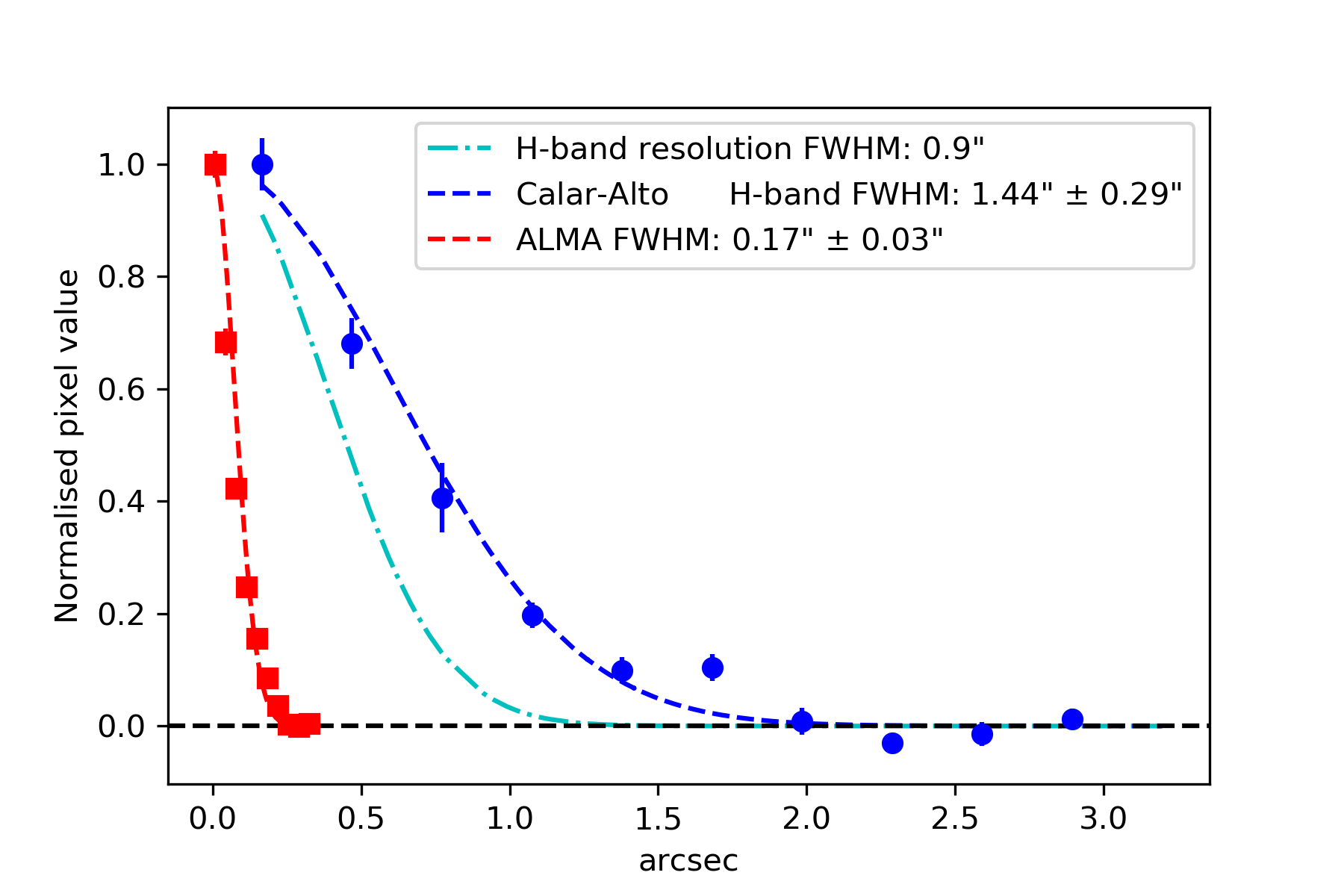}
		\caption{LAB-05}
		\label{fig:Lab5_h}
	\end{subfigure}
	\begin{subfigure}{0.495\textwidth}
		\centering
		\includegraphics[width=\textwidth]{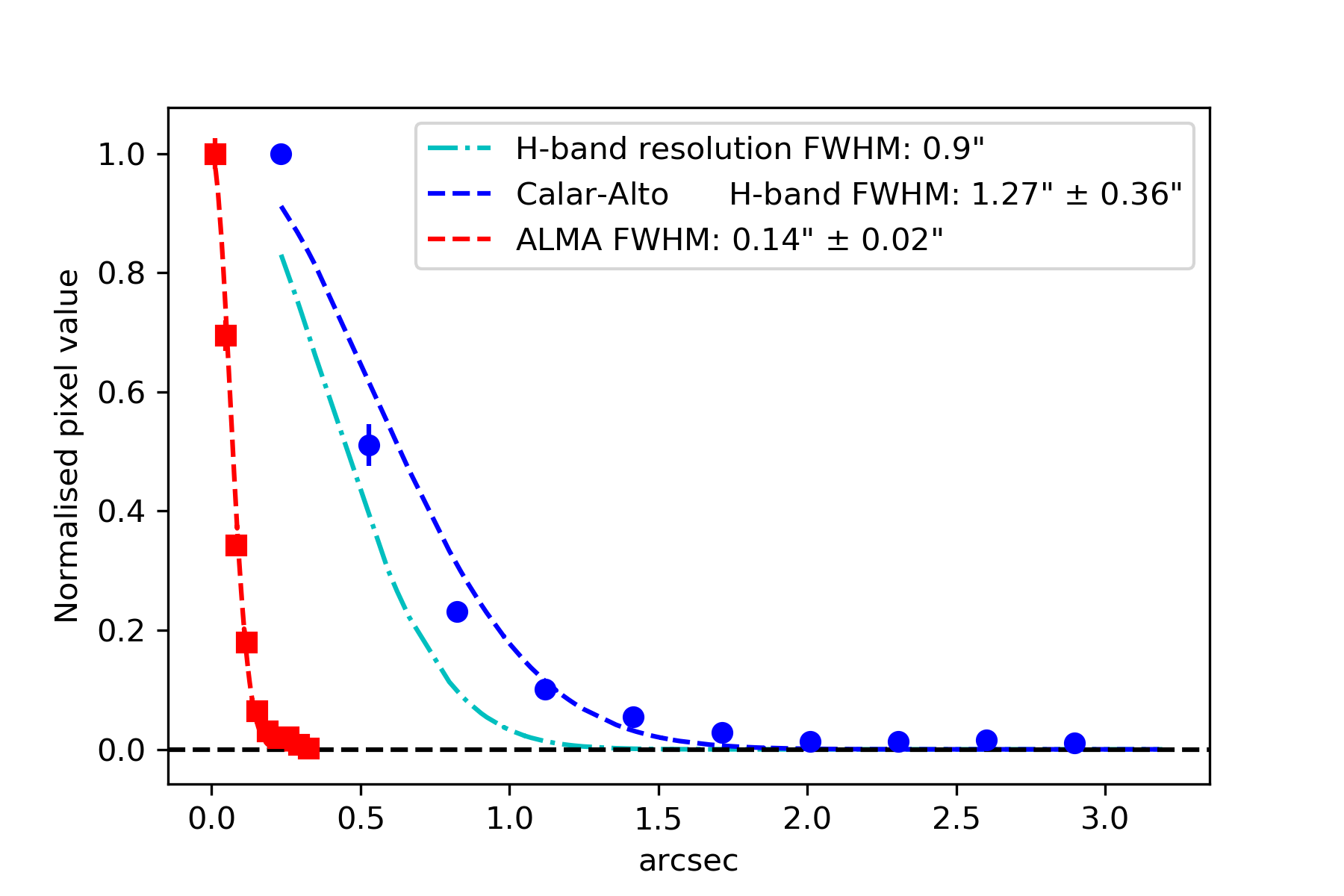}
		\caption{LAB-11}
		\label{fig:Lab11_h}
	\end{subfigure}
	\begin{subfigure}{0.495\textwidth}
		\centering
		\includegraphics[width=\textwidth]{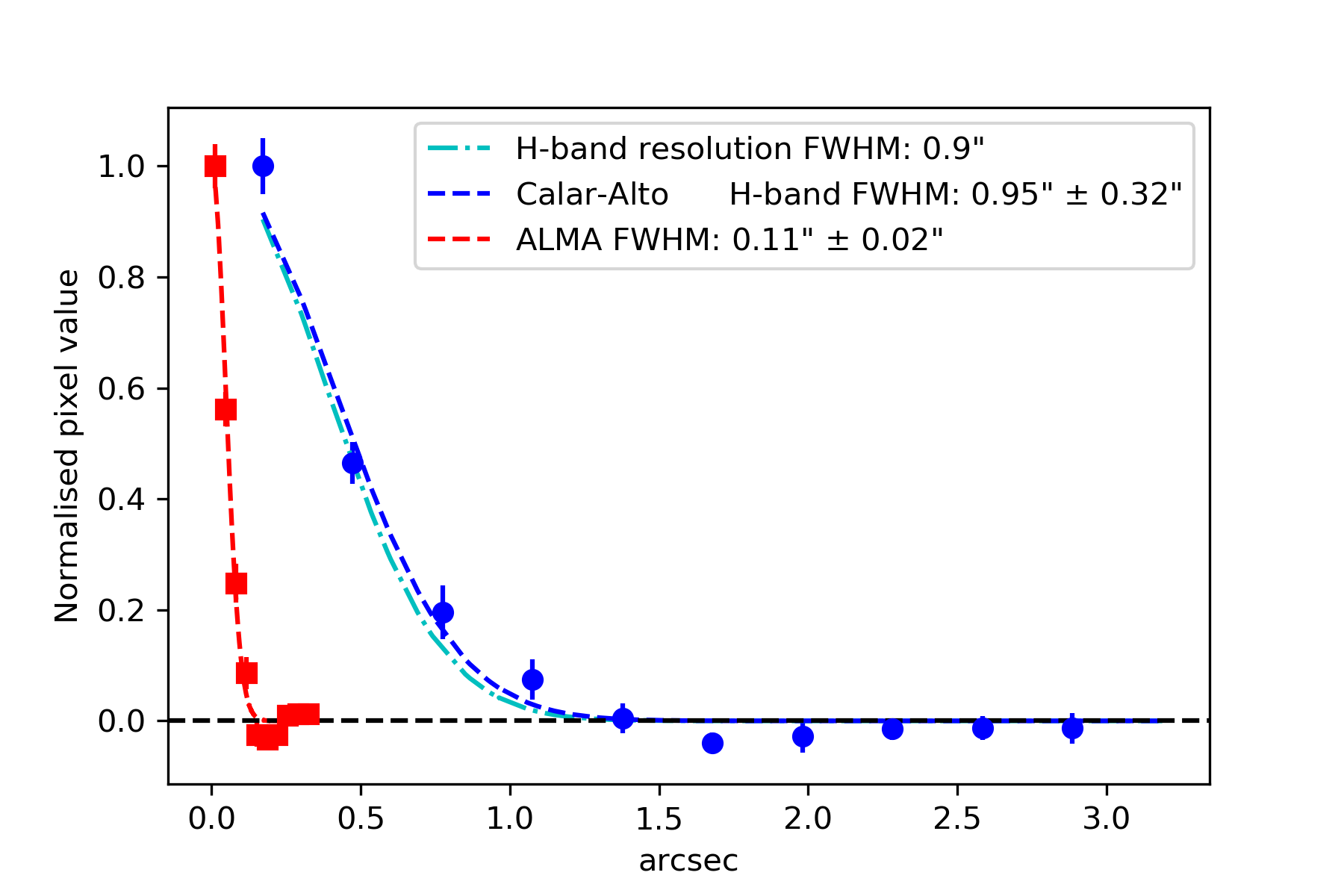}
		\caption{LAB-12}
		\label{fig:Lab12_h}
	\end{subfigure}
	\caption{Comparison of ALMA and Calar Alto $H$-band profiles of the
	WHDF sources with detections in both bands. For comparison, we
	include a Gaussian with a $0.''9$ FWHM indicating the resolution of
	the $H$-band imaging. Note that unlike in Cols. 6 and 7 of
	Table~\ref{tab:lab_size_ratios} the $H$-band source sizes in the legends
	are not corrected for the $0.''9$ seeing PSF. Similarly, no PSF correction has been applied to the ALMA profiles. The range of redshifts covered here is $1.32<z<3.0$ so the observed $H-$band corresponds to the rest
	wavelength range 4000-6900\AA~ which seems acceptably narrow. Indeed, excluding LAB-11 this range reduces still further to 4000-5100\AA, arguing that these $H$-band profiles can be consistently compared.}
	\label{fig:H-band}
\end{figure*}

\begin{table*}
	\centering
	\caption{IMFIT measurements of the WHDF SMG ALMA sub-mm sizes and fluxes.}
	\label{tab:lab_sizes}
\begin{tabular}[H]{lcccccccc}
\hline
Object & Major axis & Minor axis & Major axis  & Minor axis  & Major axis  & Minor axis  & Integrated  & Peak flux  \\
       &  FWHM      & FWHM       & FWHM\dag    & FWHM\dag    & FWHM\dag    & FWHM\dag    &flux  density&  density   \\
       &  [$''$]    &   [$''$]   &  [$''$]     &  [$''$]     & (kpc)       & (kpc)       & (mJy)       & (mJy/beam) \\
(1)    &    (2)     &    (3)     &     (4)     &    ( 5)     &     (6)     & (7)         &    (8)      & (9)  \\
\hline
LAB-01&$0.20\pm0.008$&$0.17\pm0.006$ & $0.18\pm0.010$   &$0.14\pm0.009$& $1.4\pm0.08$ & $1.1\pm0.07$& $3.84\pm0.18$ &$1.10\pm0.04$  \\[0.1cm]
LAB-02&$0.32\pm0.013$&$0.14\pm0.004$ &$0.31\pm0.014$    &$0.10\pm0.006$& $2.4\pm0.11$ & $0.8\pm0.05$& $4.68\pm0.19$ &$1.04\pm0.04$  \\[0.1cm]
LAB-03&$0.25\pm0.006$&$0.16\pm0.004$ &$0.22\pm0.007$    &$0.13\pm0.004$ & $1.8\pm0.06$ & $1.1\pm0.03$& $7.16\pm0.21$ &$1.76\pm0.04$  \\[0.1cm]
LAB-04&$0.20\pm0.024$&$0.16\pm0.017$ & $0.18\pm0.030$   &$0.12\pm0.026$ & $1.4\pm0.23$ & $1.0\pm0.20$& $1.97\pm0.3$  & $0.61\pm0.07$  \\[0.1cm]
LAB-05&$0.21\pm0.008$&$0.20\pm0.008$ & $0.18\pm0.012$   & $0.17\pm0.012$& $1.5\pm0.10$ & $1.4\pm0.10$& $5.46\pm0.26$ &$1.31\pm0.05$  \\[0.1cm]
LAB-10&$0.14\pm0.004$&$0.12\pm0.004$ & $0.094\pm0.008$  & $0.074\pm0.009$& $0.7\pm0.06$ & $0.6\pm0.07$& $2.30\pm0.10$ &$1.33\pm0.04$  \\[0.1cm]
LAB-11&$0.17\pm0.011$&$0.16\pm0.010$ & $0.14\pm0.018$   & $0.13\pm0.019$ & $1.1\pm0.15$ & $1.1\pm0.16$& $1.51\pm0.12$ &$0.55\pm0.03$  \\[0.1cm]
LAB-12&$0.20\pm0.024$&$0.10\pm0.008$ & $0.17\pm0.030$   & $0.036\pm0.021$ & $1.3\pm0.23$ & $0.3\pm0.16$& $1.68\pm0.24$ &$0.83\pm0.08$  \\[0.1cm]
\hline
\multicolumn{9}{p{0.9\textwidth}}{Notes. (1) Source short name, (2)
Major axis FWHM of the continuum emitting region, (3) Minor axis FWHM,
(4) Major axis FWHM of the continuum emitting region deconvolved from
beam, (5) Minor axis deconvolved FWHM, (6) Major axis deconvolved FWHM
(kpc), (7) Minor axis deconvolved FWHM (kpc), (8) {\sc imfit} integrated
flux densities, (9) {\sc imfit} peak flux densities. \dag~ represents extents
deconvolved from beam/PSF.}
\end{tabular}
\end{table*}

In Fig.~\ref{fig:HST_i} we compare the ALMA profiles of the WHDF sources
with the profiles of their counterpart detections from a $0.''1$
resolution HST Advanced Camera for Surveys (ACS) $i$-band imaging.
Fig.~\ref{fig:H-band} shows a comparison of the ALMA profiles of the
WHDF sources with their counterpart profiles from the H-band imaging of
\cite{Metcalfe2006} with a seeing of $0.''9$, obtained with the Calar
Alto Omega Prime camera \citep{Bizenberger1998}. The measured ALMA, $i$
and $H$-band profiles presented in these figures are all obtained using
{the {\sc iraf (v2.16.1)} {\sc
imexamine}}\footnote{\url{https://imexam.readthedocs.io/en/latest/imexam
/ iraf_imexam.html}} routine and we fit these profiles with Gaussian
functions using {a non-linear least squares fitting
technique with the {\sc scipy (v0.17.0}; \citealt{Virtanen2020})
`curve\_fit'}\footnote{\url{https://docs.scipy.org/doc/scipy/reference/
generated/scipy.optimize.curve_fit.html}} module. In
Table~\ref{tab:lab_sizes} we also list the HST $i$-band and
PSF-corrected, Calar Alto $H$-band FWHM of Gaussian profiles of the WHDF
SMGs.

Next, we describe the results for each target in turn, particularly
featuring their sub-mm + optical extents and morphologies, complementing
the treatment of \citetalias{Shanks_SMGs} where the ALMA observations were
mainly used to identify multi-wavelength counterparts to facilitate SED fitting:

\noindent{\bf LAB-01} \footnote{the SIMBAD IDs for all these objects follow the "[BHM2012] WHDF-LAB-01" SIMBAD ID naming convention.}: This source has a HST-i+SPIES MIR counterpart
classed in \citetalias{Shanks_SMGs} as a quasar at $z=2.6$ based on detection
of weak X-rays and the optical-MIR SED fit. The SED fit also gave
$A_V=1.75\pm0.25$mag and dust temperature, $T=40\pm6K$. Figs.
\ref{fig:HST_i}, \ref{fig:H-band} indicate that the counterpart is
classed as a galaxy in  HST-i with FWHM 1.8kpc and unresolved in
ground-based H band. The ALMA image in Fig. \ref{fig:lab_sources}(a) is
reasonably symmetric with deconvolved image FWHM axis sizes of
$1.4\times1.1$kpc. LAB-01 thus has an $\approx1.47\times$ larger extent
in the $i$-band compared to its ALMA FIR continuum extent. The 870$\micro$m
contours show no other low S/N features.

\noindent{\bf LAB-02}: The optical counterpart to this source was classed
as being consistent with a quasar at $z=3.1\pm0.25$ on the basis of the
SED fit with $A_V=0.25\pm0.38$ mag and $T=39\pm9K$. Fig. \ref{fig:HST_i}
shows that the counterpart is resolved with FWHM=1.5kpc. This compares
to the elongated $2.3\times0.8$kpc of the sub-mm source which lies at
$\approx7$ kpc NE of the HST $i$ source. There is also an 8.4 GHz radio
source at $\approx22$kpc from the sub-mm source. Clearly, in the sub-mm
we are seeing a dusty disc that may be small enough to be classed as a
bar, {possibly similar to those seen by \cite{Hodge2019} in
some ALESS sources, although other interpretations cannot be ruled out
e.g. a merger or a lensed galaxy.} But again no spiral features are
visible. This may be consistent with dust heating from the nucleus
affecting features at a few kpc radius. No X-ray emission was detected
at the sub-mm source position {to a limit of $\approx2\times
10^{-17}$ ergs cm$^{-2}$ s$^{-1}$ in the 0.5-2keV band which  could
still be consistent with the presence of an AGN absorbed by an edge-on
disc.} Indeed, no optical, NIR nor MIR flux was detected at the sub-mm
position and all may be similarly absorbed. This might suggest
that the HST $i$ source at $\approx7$kpc is a companion rather than {the
counterpart, as assumed here,} and this has to be borne in mind in noting that the HST $i$
and ALMA images show similar sizes in Fig.~\ref{fig:HST_i}. The actual
sub-mm source may be obscured by much more than the $A_V=0.25$ mag
estimated for the source; a Compton thick source could be obscured by
$A_V>1000$ mag and explain the lack of X-ray and even 4.6 $\micro$m
detection.

\noindent{\bf LAB-03}: With just a marginal detection in the $H$-band at
the ALMA position, no counterpart was claimed for this SMG, only a
companion  at $1.''82$ distance from the sub-mm source. The SED fit for
the companion implied $z=2.7\pm0.35$ and assuming both are at the same
redshift, they are separated on the sky by  $\approx15$kpc. X-ray
emission was detected at  the SMG position, {corresponding
to $L_\textup{X}(1.2-2${keV}$)\approx1.1\times10^{43}$erg s$^{-1}$,} and it was this
that was crucial in its identification as a probable quasar\footnote{Fig. 16 of \citet{Luo2017} shows that in their analysis 
of the Chandra 7 MSec observations of Chandra Deep Field-South that 
$L_{\rm X}(0.5-7keV)$<$3\times10^{42}$ergs s$^{-1}$ is already a conservative
upper limit on the X-ray luminosity of a star-forming galaxy due to
Low-Mass X-ray Binaries, with very few examples of star-forming galaxies
being seen above the usual quasar limit of $L_{\rm X}>1\times10^{42}$ergs
s$^{-1}$ (e.g. \citealt{Zezas1998, Moran1999}).}. The sub-mm
source has elongation between that of LAB-01 and LAB-02 with extent
1.8kpc$\times1.1$kpc. Again the difficulty in finding an optical or MIR
counterpart could be due to significant dust absorption and this might
be related to being seen at an angle rather than face-on. No low surface
brightness sub-mm features were detected.

\noindent{\bf LAB-04}: Although SED fitting marginally preferred a
star-forming galaxy template, the counterpart was  detected in  X-rays,
{with a luminosity corresponding to  $L_\textup{X}(1.2-2$
keV$)\approx1.8\times10^{43}$erg s$^{-1}$ at $z\approx3$,} and its red
[3.6]-[4.5] colour was also consistent with it being a quasar. The AGN
SED fit gave $z=3.0$ and $A_V=2.5$mag, in line with its lack of UBRI
detection. At this redshift, its deconvolved extent is 1.4 kpc $\times$
1.0 kpc so again its sub-mm morphology is compact and reasonably
symmetric. LAB-04 has a detection in the $H$-band with an
$\approx16\times$ larger FWHM extent ($\approx18$ kpc) relative to the
ALMA profile of this source ($\approx1.2$ kpc, see
Table~\ref{tab:lab_size_ratios} and Fig.~\ref{fig:H-band}).

\noindent{\bf LAB-05}: This object is identified as an X-ray absorbed
quasar at a spectroscopic redshift of $z=2.12$, and SED fitting gave
$A_V=1.0$ mag. Table \ref{tab:lab_sizes} shows ALMA extents of 1.4 kpc
$\times$ 1.2 kpc and so this source is barely resolved. The HST $i$
image in Fig. 1 of \citetalias{Shanks_SMGs} shows that LAB-05 has a
blotchy appearance with several possible nuclei. Despite the dust
absorption, the quasar shows a strong ultra-violet excess with
(U-B)$_{\rm Vega}=-1.25$ mag \citep{Heywood2013}. The absorbed  X-ray source
and the sub-mm source are reasonably coincident with one of these nuclei
that is slightly offset with respect to the $i$-band image. We see that
the size of the $i$-band nucleus is similar in size to the sub-mm image
at $\approx 1.0-1.4$ kpc. But the full extent of the host galaxy is
larger ($\approx2.7\times$) and resolved even on the $0.''9$ scale of
the $H$-band image with a deconvolved extent of $0.''5$ or $\approx4$
kpc. Thus the sub-mm source seems highly compact and associated directly
with the AGN. This looks similar to what is seen in SMGs LAB-01, LAB-02
and LAB-04.

\noindent{\bf LAB-10}: This sub-mm source also showed no direct
counterpart in any band other than 250 and 350$\micro$m and its redshift of
$z=0.9$ was estimated from a companion at $\approx9$ kpc separation.
Assuming this redshift, LAB-10 is $\approx0.7$ kpc in extent (see
Table~\ref{tab:lab_size_ratios}) and again highly compact and barely
resolved with no evidence of low surface brightness sub-mm features (see
Fig.~\ref{fig:ALESS_sources_ALMA}). Otherwise, with no counterpart no
further conclusion can be drawn except that the host must be highly
obscured.

\noindent{\bf LAB-11}: This sub-mm source is identified with a $z=1.32$
X-ray absorbed quasar and SED fitting gives $A_V=1.5\pm0.25$ mag and
dust temperature $T=41\pm10$K. The X-ray source is coincident with the
sub-mm source. However, the quasar X-ray absorption is much higher than
implied by the fitted dust absorption. Despite the dust absorption, the
quasar still shows some ultra-violet excess with (U-B)$_{\rm Vega}=-0.72$ mag
\citep{Heywood2013}. There is also an 8.4 GHz radio source at $1.''2$
from the ALMA position. 

The optical structure of the LAB-11 host galaxy is clearly resolved at
HST $i-$ band resolution with an extent of $\approx3$ kpc,
$\approx3\times$ larger than the 1.1 kpc $\times1.1$ kpc of  the sub-mm
source which is barely resolved at $0.''1$ ALMA resolution. Fig. 1 of
\citetalias{Shanks_SMGs} shows that the host galaxy of LAB-11 has a smoother,
less nucleated structure than the LAB-05 quasar host.  Fig.
\ref{fig:HST_i} shows that LAB-11 is clearly detected in the $i$-band
where it has an extent $\approx3\times$ larger than its counterpart ALMA
detection, while it is only marginally resolved in the $H$-band (see Fig.
\ref {fig:H-band} and Table \ref{tab:lab_size_ratios}).

\noindent{\bf LAB-12}: This sub-mm source was detected for the first time
in the ALMA observation at $10.''6$ from LAB-11 and is not in the flux
limited sample. Nevertheless, an optical+MIR counterpart was detected
and SED fitting and MIR colour were consistent with it being a probable
quasar at $z=2.9$. Its sub-mm structure appears elongated with extents
of 1.3 kpc $\times$ 0.3 kpc and Fig. 1 of \citetalias{Shanks_SMGs} shows that the
HST $i-$band counterpart has a similar elongation but with $\approx2\times$
larger scale (see also Fig. \ref{fig:HST_i} and Table
\ref{tab:lab_size_ratios}). So  in this lower luminosity sub-mm source, the
sub-mm and optical morphologies and scale look more similar to what
naively might be expected if the dust was heated  more by {\it in situ}
star-formation.

\begin{table*}
	\centering
	\caption{IMFIT+IRAF measurements of the WHDF SMG sizes in sub-mm, $i$ and $H$ bands and their ratios.}
	\label{tab:lab_size_ratios}
\begin{tabular}{lcccccc|cc}
\hline
Object & FIR Major   & FIR Minor   & $i$-band    &   $H$-band     & $i$-band       & $H$-band      & i-band/FIR  &	$H$-band/FIR\\
       & FWHM\dag    & FWHM\dag    & FWHM        &   FWHM       & FWHM\dag       & FWHM\dag    &  ratio      &   ratio  \\		
\hline
LAB-01& $0.18\pm0.010$& $0.14\pm0.009$& $0.23\pm0.03$&$1.01\pm0.52$&$0.21\pm0.020$&$0.46\pm0.52$&$1.34\pm0.13$&$2.92\pm1.14$ \\[0.1cm]
LAB-02& $0.31\pm0.014$& $0.10\pm0.006$ & $0.22\pm0.03$&    ---      &$0.20\pm0.020$&     ---     &$0.98\pm0.11$&    ---      \\[0.1cm]
LAB-03 &$0.22\pm0.007$& $0.13\pm0.004$ &    ---      &    ---      &     ---        &     ---     &    ---      &    ---      \\[0.1cm]
LAB-04 & $0.18\pm0.030$& $0.12\pm0.026$&    ---      & $2.57\pm1.22$&     ---        &$2.40\pm1.22$&    ---      &$16.0\pm8.66$\\[0.1cm]
LAB-05& $0.18\pm0.012$& $0.17\pm0.012$ & $0.19\pm0.011$ & $0.14\pm0.029$ &   $0.16\pm0.11$  &$1.12\pm0.29$ &$0.92\pm0.63$&$6.35\pm1.69$*\\[0.1cm]
LAB-10 & $0.094\pm0.008$  & $0.074\pm0.009$      &    ---      &    ---       &     ---        &    ---      &    ---      &    ---      \\[0.1cm]
LAB-11 & $0.14\pm0.018$& $0.13\pm0.019$    & $0.38\pm0.060$  & $1.27\pm0.36$ &   $0.37\pm0.060$   & $0.90\pm0.36$ &$2.80\pm0.60$&$6.79\pm2.89$\\[0.1cm]
LAB-12 & $0.17\pm0.030$& $0.036\pm0.021$     & $0.24\pm0.15$ & $0.95\pm0.32$  &   $0.22\pm0.15$  & $0.30\pm0.32$ &$2.16\pm1.59$&$3.01\pm3.26$\\[0.1cm]
\hline
\multicolumn{9}{p{0.9\textwidth}}{Notes. All columns except the last two are in  units of arcseconds.
\dag~ represents extents deconvolved from beam/PSF. *LAB-5 is
associated with a multiply nucleated host galaxy which is resolved into
separate objects in the $i$-band but not the $H$-band. {In this case, the} {$i$-band size is measured from the nucleus nearest the SMG and the $H$-band size is measured for the entire host galaxy.}}
\end{tabular}
\end{table*}

\subsection{Comparison with  ALESS SMG sizes and morphologies}
\label{sec:aless}

\begin{table*}
	\caption{Noise levels (standard deviations) measured in $\micro$Jy/beam in vicinity of  ALMA targets.}
	\label{tab:noise}
\begin{tabular}[H]{lccccccccccc}
\hline
LAB-01   &LAB-02   &LAB-03   &LAB-04   &LAB-05   &LAB-10   &LAB-11   &LAB-12   & J332-23   &J158-14  &J025-33   &J029-36\\
$\pm36.8$&$\pm38.6$&$\pm36.7$&$\pm68.8$&$\pm36.7$&$\pm35.3$&$\pm35.9$&$\pm79.7$& $\pm25.8$ &$\pm24.7$&$\pm29.9$ &$\pm31.4$\\
\hline
\end{tabular}
\end{table*}

For comparison, in Fig.~\ref{fig:ALESS_sources_ALMA} we plot high
resolution ($\approx0.''07$) ALMA observations of the six luminous SMGs
from the ALESS sample  \citep{Hodge2013,Hodge2016} studied by
\cite{Hodge2019}. { We chose to compare to these data
because their combination of ALMA  high resolution and  exposure time is
more comparable to our  WHDF data than for other samples such as those
of  e.g. \citet{Ikarashi2017}, \citet{Elbaz2018} or
\citet{Gullberg2019}. The ALESS flux density range also overlaps our
WHDF range more than the fainter sub-mm galaxy samples from
\citet{Rujopakarn2016,Rujopakarn2019}, \citet{Franco2020} or
\citet{GomezG2021}.} Fig.~\ref{fig:ALESS_sources_ALMA} shows that unlike
the WHDF SMGs, the ALMA observations of ALESS SMGs reveal potential
evidence of sub-structure detection in these sources. Furthermore, the
ALESS SMGs appear to have half-light radii\footnote{The half-light
radius of a source is defined as the mean of its major and minor axes'
FWHM/2.} that are on average $\approx3\times$ larger than
those of the WHDF SMGs. {The average ALESS SMG flux density is  also
$\approx2\times$ brighter than that of the WHDF SMGs.} Given the factor of
$\approx2$ lower exposure time and lower resolution ($0.''1$ vs
$0.''07$) of the ALMA observations of the WHDF sources compared to those
of the ALESS SMGs, it might be thought that sub-structure currently remains
undetected in the WHDF observations. {However, the average
surface brightness of $\approx200\micro$Jy/beam is clearly reached by all
the WHDF sub-mm observations (see Table \ref{tab:noise}). The two highest noise
cases are LAB-04 and LAB-12 with $\pm70-80 \micro$Jy/beam and even they
would detect the ALESS diffuse structure at $2.5-2.8\sigma$ over just
one beam size and clearly more if averaged over a large area. These
significances of detection would rise to $\approx5-6\sigma$ for the
other WHDF SMGs.}

Another possibility, however, is that despite the fact that both samples
occupy a similar redshift range, there is an inherent difference between
the two SMG populations. While the WHDF SMGs were initially detected in
blind LABOCA observation of the WHDF, the ALESS SMGs studied by
\cite{Hodge2019} were selected as the brightest amongst the large ALESS
sample (see \citealt{Hodge2013}; \citealt{Hodge2016}). This selection
could in turn bias the sample towards selecting larger than average SMGs
that are more likely to contain distinct sub-structures. {This would also be 
consistent with previous claims of a slow increase in SMG FIR size with FIR luminosity (e.g. \citealt{Fujimoto2017, Gullberg2019}). }

\subsection{Comparison with $z>6$ quasar sizes and morphologies}
\label{sec:z6quasarsize}

We also analysed our four $z>6$ quasars using {\sc imfit} in the same way 
as for the WHDF SMGs. The resulting sub-mm images are shown in 
Fig.~\ref{fig:hiz_sources_ALMA} and their sizes in Table \ref{tab:z6_sizes}.
With the exception of J025-33,  all our other sources
appear to be resolved. We now discuss each quasar individually. 

\noindent{\bf J332-23} The sub-mm counterpart of this $z=6.32$ quasar is
just  resolved by ALMA at $0.''4\times0.''28$ resolution. It has an 
apparently elongated sub-mm morphology with deconvolved extents measured
as 0.9$\times0.5$kpc. However, the beam is also elongated for this
observation and this means large errors on these deconvolved sizes. The
longer deconvolved axis is detected at $2.5\sigma$ and the shorter axis
is detected at $1.7\sigma$. However, the evidence against circularity is
only significant at $0.8\sigma$. We conclude that this image is resolved
but with only marginal evidence for elongation. We note that a second
unidentified sub-mm source, with integrated and peak flux densities $0.66\pm0.055$ mJy  
and $0.45\pm0.024$ mJy, is detected by ALMA at $5.''4$ SW from the
source associated with the quasar.

\noindent{\bf J158-14} The sub-mm counterpart of this  $z=6.07$ quasar
is again indicated as resolved by {\sc imfit} with deconvolved extents
of 1.1 kpc $\times$ 0.8 kpc. Again there is no indication against a
symmetric, circular sub-mm image, given the errors on these extents
quoted in Table \ref{tab:z6_sizes}.

\noindent{\bf J025-33}  at $z=6.31$ was identified as a point source in
the sub-mm by {\sc imfit} with a FWHM major axis $<0.''2\times0.''07$
giving an upper limit in physical size of 1.1kpc$\times0.4$kpc. In
Fig.~\ref{fig:hiz_sources_ALMA} (b), J025-33 shows a relatively circular
shape, similar to the beam with its FWHM of  $0.''38\times0.''34$ and
{\sc imfit} only gives an upper-limit ($<0.''2$ or $<1.0$ kpc) on the
extent of this source. Consequently, in Section~\ref{sec:SMG_results} we
only present a lower-limit for the star formation rate surface density
($\sum_{\textup{SFR}}$) of this object. We note that J025-33 lies closer
to the edge of the field in the ALMA observations and therefore has a
lower S/N compared to the other detections, which lie closer to the
field centre.

\noindent{\bf J029-36} This is the most clearly resolved  quasar with
deconvolved sky extents of $0.''33\pm0.''02\times0.''22\pm0.''02$
which at $z=6.02$ translate to 1.9 kpc $\times$ 1.2 kpc. Again there is
little evidence of non-circularity in the sub-mm image in
Fig.~\ref{fig:hiz_sources_ALMA} or in Table \ref{tab:z6_sizes}.

Thus, helped by the angular diameter distance-redshift relation,
$d_\textup{A}(z)$, we can resolve 3 out of 4 $z>6$ quasar images despite the
lower $\approx0.''33$ ALMA resolution than for the WHDF SMGs. Assuming
the upper limit for the unresolved extent of the J025-33 host we find an
average of $1.25\pm0.22$ kpc for the average sub-mm size of the major
axes of the quasar hosts in the sub-mm. This compares with
$1.44\pm0.17$kpc for the average of the major axes of the 8 WHDF SMGs
given in Table \ref{tab:lab_sizes} which is statistically consistent
with the $z>6$ quasar sizes after beam deconvolution.

{The high redshift of the ATLAS $z>6$ QSOs makes any search
for low surface brightness components more difficult than for the WHDF
SMGs. Since there is typically a factor of $\approx2$ difference in redshift the
$(1+z)^4$ dimming law means that they will be $\approx10\times$ lower or
$\approx20\micro$Jy. An average of $\approx100$ beam areas (i.e. only a
$3''\times3''$ extent) would be needed to detect them but none are seen.
Although this result appears  contrary to the FIR size-luminosity
relation mentioned in Section \ref{sec:aless}, this might be due to the
significantly higher redshift of these sources. At minimum, this means that
our search for a diffuse component is at $\approx$150$\micro$m rest
wavelength rather than  $\approx$300$\micro$m for the ALESS/WHDF SMGs.}

\begin{table*}
	\centering
	\caption{IMFIT measurements of the $z>6$ quasar ALMA FIR source sizes and fluxes.}
    \label{tab:z6_sizes}
	\begin{tabular}[H]{lccccccc}
		\hline
		 Object & $\lambda$ & Beam size & Major axis  & Minor axis & Area & Integrated flux & Peak flux \\
		        &           &           & FWHM        & FWHM       &      &  density        &  density  \\
                & ($\micro$m)  &[$''$]     & [$''$]      & [$''$]     &(kpc$^2$)& (mJy)        & (mJy/beam) \\
         (1)    & (2)       & (3)       & (4)         & (5)        & (6)   & (7)            &  (8) \\
\hline
J332-23     & 1202.27 & $0.40\times0.28$ & $0.16\pm0.07$ & $0.09\pm0.05$ & $0.3\pm0.2$ & $0.47\pm0.03$ & $0.41\pm0.02$ \\[0.1cm]
J158-14     & 1150.06 & $0.36\times0.29$ & $0.20\pm0.02$ & $0.14\pm0.02$ & $0.7\pm0.1$ & $3.21\pm0.07$ & $2.50\pm0.03$ \\[0.1cm]
J025-33     & 1192.47 & $0.38\times0.34$ & $<0.2$        & $<0.2$        &  $<1.0$     & $0.76\pm0.05$ & $0.70\pm0.03$ \\[0.1cm]
J029-36     & 1145.16 & $0.34\times0.31$ & $0.33\pm0.02$ & $0.22\pm0.02$ & $1.8\pm0.2$ & $1.99\pm0.09$ & $1.17\pm0.03$  \\[0.1cm]
\hline
\multicolumn{8}{p{0.8\textwidth}}{Notes. (1) Source short name, (2) reference wavelength, (3) ALMA
clean beam size, (4) Major axis FWHM of the continuum source
deconvolved from beam, (5) Minor axis deconvolved FWHM, (6) Surface area of
continuum emitting region, (7) {\sc imfit} integrated fluxes, 
(8) {\sc imfit} peak fluxes.
}
	\end{tabular}
\end{table*}

\section{SMG and quasar FIR luminosities and SFR  densities}
\label{sec:SMG_SFRs}

Following \citet{Decarli2018} and \cite{Beelen2006}, we calculate the FIR
Luminosity of our SMGs and quasars by modelling the dust continuum
emission as a modified black body. Specifically, we estimate  $L_{\textup{FIR}}$
to the observed $S_{870 \textnormal{\rm \micro m}}$ (WHDF SMGs) and $S_{1200 \textnormal{\rm \micro m}}$
($z>6$ quasars) flux density using  eqs (1) and (4) of \cite{Beelen2006}.
Assuming dust  temperature $T_{\textup{dust}}$ and with $\nu_\textup{r}$ the rest
frequency corresponding to e.g. $870\micro$m at redshift, $z$, we calculate
$M_{\textup{dust}}$ from:

\begin{equation}
S_{870,\textup{obs}}=\frac{1+z}{d_\textup{L}^2}  {M_{\textup{dust}}} B_{\nu_\textup{r}}(T_{\textup{dust}}) \kappa(\nu_\textup{r}) 
\label{eq:beelen1}
\end{equation}

\noindent where

\begin{equation}
    B_{\nu}(T_{\textup{dust}})=\frac{2h\nu^3}{c^2}\frac{1}{e^{h\nu/k_b T_{\textup{dust}}}-1},
	\label{eq:beelen2}		
\end{equation}

\noindent is the Planck function and 
$\kappa(\nu)=0.077(\nu/352\textup{GHz})^\beta$m$^2$kg$^{-2}$ is the
opacity law, with dust emissivity index, $\beta=1.6$ {(\citealt{Beelen2006,Decarli2018})}. We then use
this value for  $M_{\textup{dust}}$  to calculate $L_{\textup{IR}}$ from:

\begin{equation}
    L_{\textup{IR}}=4\pi\, M_{\textup{dust}} \int{ B_\nu(T_{\textup{dust}}) \kappa(\nu)  d\nu}
	\label{eq:beelen3}
\end{equation}

\noindent with the integral over frequencies, $\nu$, corresponding to
rest wavelengths between ${\rm 3-1100 \micro}$m (following
\citealt{Kennicutt2012,Decarli2018}). In this work, we set the dust
temperature to the AGN dust temperature estimated in
\citetalias{Shanks_SMGs} when available and otherwise
$T_{\textup{dust}}=35$K in the case of our WHDF sources (e.g.
\citealt{Cooke2018}). We take  $T_{\textup{dust}}=47$K for our $z>6$
quasars (see e.g. \citealt{Willott2017}). We then obtain the SFR
following \cite{Kennicutt2012}:

\begin{equation}
    \frac{\textup{SFR}_{\textup{IR}}}{\textup{M}_\odot \textup{yr}^{-1}}=1.49\times10^{-10}\frac{L_{\textup{IR}}}{\textup{L}_\odot}.
	\label{eq:beelen4}
\end{equation}

Following \cite{Hodge2019}, we report two estimates of the star formation
rate surface density, the integral or galaxy-averaged
$\sum_{\textup{SFR}}$ and its peak value. The galaxy-averaged
$\sum_{\textup{SFR}}$ is given by $(0.5\times
\textup{SFR}_{\textup{IR}})/(\pi R_e^2)$ where $R_e$ is the half-light
radius {defined as in Section \ref{sec:aless} as the mean of its major and minor axes'
FWHM/2.}
The peak $\textup{SFR}$ density  is calculated using the
SFR$_\textup{IR}$ value based on the  peak flux density/beam and divided by the
beam area defined from its major and minor axes at FWHM, $a, b$, as
$\frac{\pi ab}{4\textup{ln}(2)}$. {Clearly, the closer a sub-mm source
is to being unresolved the less difference there will be between the integrated 
and peak values and we shall see that this applies to many of the WHDF SMGs and $z>6$
quasars considered here.}

\begin{table*}
	\centering
	\caption[SMG Properties]{SMG and quasar properties estimated  from ALMA FIR/IR continuum flux densities and FIR sizes.}
	\label{tab:SFR}
		\begin{tabular}[H]{lcccc|c} 
		\hline
		 Object & $M_{\textup{dust}}$ & $L_{\textup{FIR}}$ & $\textup{SFR}_{\textup{IR}}$ & Galaxy-averaged $\sum_{\textup{SFR}}$ & Peak $\sum_{\textup{SFR}}$ \\
        & ({$10^8\textup{M}_\odot$}) & ({$10^{12}\textup{L}_\odot$}) & $(\textup{M}_\odot \textup{yr}^{-1}$) & ($\textup{M}_\odot$yr$^{-1}$kpc$^{-2}$) &  ($\textup{M}_\odot$yr$^{-1}$kpc$^{-2}$)  \\
        (1) & (2) & (3) & (4) & (5) & (6)  \\
		\hline
        LAB-01 & {$5.0\pm1.0$} & {$3.4\pm1.8$} & {$663\pm359$} & {$257\pm146$}& {$265\pm143$} \\[0.1cm]
        LAB-02 & {$6.1\pm1.7$} & {$3.7\pm3.0$} & {$707\pm586$}     & {$184\pm155$} & {$240\pm199$} \\[0.1cm]
        LAB-03 & {$12.0\pm3.6$} & {$3.9\pm2.2$} & {$760\pm437$} & {$251\pm146$} & {$264\pm152$} \\[0.1cm]
        LAB-04 & {$3.3\pm1.5$} & {$1.1\pm0.7$} & {$206\pm142$} & {$98\pm86$}  & {$95\pm65$}   \\[0.1cm]
        LAB-05 & {$8.9\pm1.3$} & {$3.4\pm1.7$} & {$659\pm323$}     & {$198\pm104$} & {$204\pm100$} \\[0.1cm]
        LAB-10 & {$7.9\pm4.6$} & {$<0.18^*$}   & {$<68^*$}         & {$<103^*$}   & {$<58^*$}    \\[0.1cm]
        LAB-11 & {$1.9\pm0.5$} & {$1.5\pm1.3$} & {$301\pm258$}     & {$149\pm140$}& {$390\pm333$} \\[0.1cm]
        LAB-12 & {$2.8\pm0.9$} & {$0.9\pm0.5$} & {$177\pm93$}      & {$175\pm155$}& {$128\pm66$}  \\[0.1cm]        
		\hline
	    J332-23 & {$0.65\pm0.1$} & {$1.0\pm0.4$} & {$215\pm81$}     & {$<679^*$}   & {$48\pm18$}   \\[0.1cm]
		J158-14 & {$4.1\pm0.4$} & {$6.5\pm2.4$} & {$1360\pm506$}   & {$943\pm472$} & {$277\pm103$} \\[0.1cm]
		J025-33 & {$1.0\pm0.1$} & {$1.6\pm0.6$} & {$344\pm130$}    & {$>387^\dagger$}     & {$70\pm26$}   \\[0.1cm]
        J029-36 & {$2.5\pm0.3$} & {$4.0\pm1.5$} & {$837\pm314$}    & {$219\pm98$}  & {$126\pm47$} \\[0.1cm]
		\hline
        \multicolumn{6}{p{0.8\textwidth}}{Notes. (1) Source short name,
        (2) Dust mass; these estimates are to be preferred over 
        those quoted in \citetalias{Shanks_SMGs}, {(3) FIR luminosity calculated using eq. \ref{eq:beelen3}
        with rest wavelength limits between $42.5-122.5\micro$m},
        (4) {IR }star formation rate,
        (5) {IR }galaxy-averaged star formation rate surface density ($^\dagger$ signifies lower 
        limit when maximum area of 1 kpc$^2$ for J025-33 is assumed),
        (6) {IR }peak star formation rate surface 
        density (see text for details). {In columns 3, 4, 5, 6 for LAB-10, $^*$ signifies a $1\sigma$ upper limit.}}
	\end{tabular}
\end{table*}

\subsection{SFR densities and  the SFR `Eddington limit'}
\label{sec:SMG_sfr}

The results of fitting the continuum emission  maps of our sources using the {\sc imfit}
algorithm, are presented in Tables~\ref{tab:lab_sizes} and
\ref{tab:z6_sizes}. The clean beam size for all WHDF observations was
$0.''11\times0.''09$ while those for the $z>6$ quasars vary slightly as
shown in Table \ref{tab:z6_sizes}. Also shown in both cases, are the
beam deconvolved minor and major axes FWHM of the fits which are used to
estimate the area of the continuum  emitting regions. Also included are the 
integrated and peak flux densities of each source as measured by {\sc imfit}.  

Table~\ref{tab:SFR} shows the dust mass, FIR luminosities, star formation rates and
star formation rate surface densities of the WHDF and ATLAS $z\approx6$
SMGs. We emphasise that the errors on these quantities may be
underestimated because we have not included systematic errors such as
the variation of the dust emissivity index, $\beta$, which may vary within sources
as well as between sources. Similarly, dust temperatures are also assumed
to be constant within individual sources and again this may turn out to
be an over-simplification. Thus the absolute values of quantities such
as the inferred dust mass for individual sources may  be less reliable
than inferred from the quoted statistical error. We should be on safer ground  when
we note e.g. that the dust masses for the 6 SMGs and the 6 quasars in
Table~\ref{tab:SFR} mostly lie within the narrow range $\approx10^8 - 10^9 \textup{M}_\odot$
although, even here, these assumptions have to be borne in mind. With
this proviso, we further note that the FIR luminosities and
star-formation rates also lie in the same ranges for the SMGs and quasars
in Table~\ref{tab:SFR} and we shall return to discuss these
similarities further in Section \ref{sec:SMG_discussion}.

\begin{figure}
    \centering
        \includegraphics[width=\columnwidth]{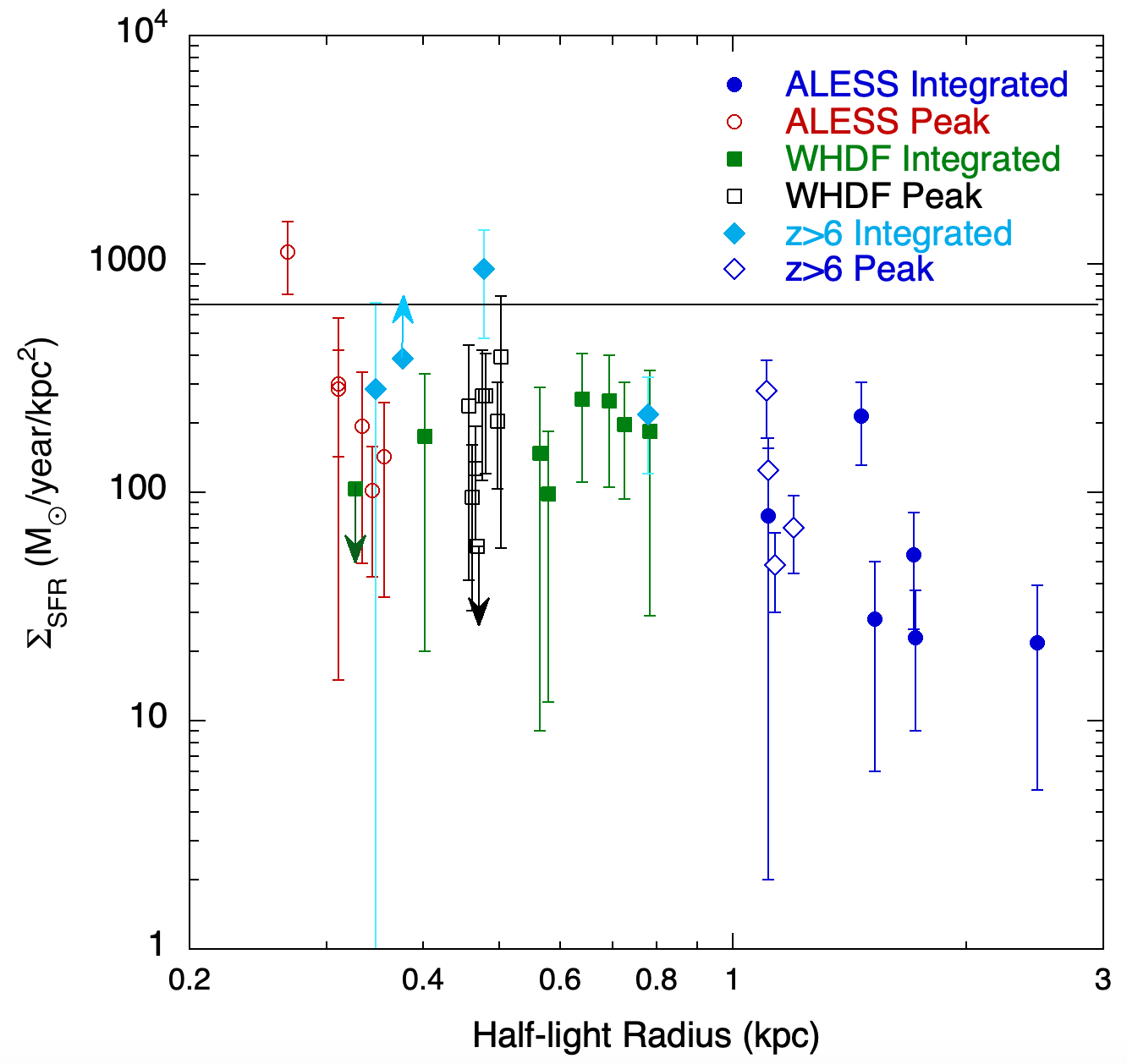}
	\caption[SFR surface densities vs. half light radii]{The integrated and peak star
	formation rate surface density ($\sum_{\textup{SFR}}$) as a function
	of half-light radius for 
	the ALESS SMGs of \cite{Hodge2019}, compared to the WHDF SMGs and the
	$z\approx6$ AGN presented in this work. The solid horizontal line
	indicates a lower bound for the Eddington limit for star formation as calculated
	by \cite{Hodge2019}.
	}
	\label{fig:Hodge}
\end{figure}

In Fig.~\ref{fig:Hodge}, we plot the integrated and peak
$\sum_{\textup{SFR}}$ values of our sources versus respectively the
effective radius, R$_e$, and average beam size, $(a+b)/2$. For
comparison, we also include  measurements of  the ALESS SMG sample, as
presented in Fig. 6 of \cite{Hodge2019} where they are compared to a
`lower bound` to the `Eddington limit' of SFR surface density, as
derived in Section 4.1 of \cite{Hodge2019}. This lower bound was
obtained by converting the Eddington flux for optically thick starbursts
(as given by \citealt{Andrews2011}) to the Eddington limited SFR density:
$(\sum_{\textup{SFR}})_{\textup{Edd}}\approx7.2 \textup{M}_\odot$yr$^{-1}$kpc$^{-2
}f_ {\textup{gas}}^{-1/2}f_{\textup{dg}}^{-1}$. Here, the gas fraction
$f_{\textup{gas}}$ is taken to be unity as the most extreme scenario,
while the dust-to-gas ratio $f_{\textup{dg}}$ is assumed to be
$1/90$, { as proposed for SMGs by
\citet{Magnelli2012,Swinbank2014}. It is this first assumption } that
makes the value they derived, $\sum_{\textup{SFR}}\approx650
\textup{M}_\odot$yr$^{-1}$kpc$^{-2}$, a lower bound for this quantity. Note that
in Fig.~\ref{fig:Hodge}, the SFR surface density values for the 6 ALESS
SMGs have been re-calculated according to eqs
\ref{eq:beelen1}-\ref{eq:beelen4} for consistency with our SMG and
quasar values. We find these recalculated values for the integrated  SFR
densities {are a factor of $2.1\pm0.23$ higher than those
given by \cite{Hodge2019} due to their different estimation method. Our 
peak SFR densities are similarly  a factor of $2.2\pm0.19$ higher.} One
result is that the peak SFR density of ALESS 9.1  now appears above the
Eddington limit. The known X-ray quasar ALESS 17.1 remains a factor of
$\approx7\times$ below this limit.


All of the WHDF sub-mm sources lie below the Eddington limit and close
to  the ALESS sources in their SFR density, with the peak and
integrated values generally being generallty similar within the errors for both the
WHDF and $z>6$ sources. Here, the two known quasars LAB-05 and LAB-11 have
similar SFR surface densities to the others. The slightly higher ALESS peak
values relative to WHDF are probably due to their $\approx30$\% higher
resolution than the WHDF data.

For the four $z>6$ quasars, we see in Fig.~\ref{fig:Hodge} that the peak
values are again slightly lower than those for ALESS and WHDF SMGs. Again
these lower peak values might be explained by the $\approx3\times$ lower
resolution in arcseconds, although given the $d_\textup{A}(z)$ relation, this is
only $\approx2\times$ lower resolution measured in kiloparsecs.

Summarising, we have compared  the integrated and peak
$\sum_{\textup{SFR}}$ values of our {six WHDF SMGs with
six} known sub-mm-loud quasars. For comparison, we also include
$\sum_{\textup{SFR}}$ measurements of
the ALESS SMG sample, as presented in Fig. 6 of \cite{Hodge2019}. We
find that peak SFR density values generally increase with spatial
resolution measured in kiloparsecs. Only one ALESS SMG exceeds the
SFR Eddington limit but none of the others do, except for some
highly resolved local U/LIRGs (see Fig. 6 of \citealt{Hodge2019}). The 5
WHDF SMGs show little difference with the 2 WHDF sub-mm quasars, the 4
$z>6$ quasars and 5 out of 6 ALESS SMGs. The only further statement that
can be made is that perhaps ALESS 9.1 with its super-Eddington SFR
density might now join the X-ray source ALESS 17.1 in making at least 2
out of the 6 sources of \cite{Hodge2019} now identified as AGN. Although
these statistics are still too poor to claim 1/3 of ALESS sources as a
lower limit to the quasar fraction, we do note that ALESS 9.1 and 17.1
share similar low surface brightness galaxy bar and spiral arm features
as the four other ALESS sources. So in interpreting these results, the
morphological dissimilarity between ALESS and WHDF SMG samples will have
to be explained by some more general sample characteristic e.g. FIR
luminosity or S/N, rather than by sample bias towards one or other class
of heating source.

\begin{figure*}
    \centering
	\includegraphics[width=\textwidth]{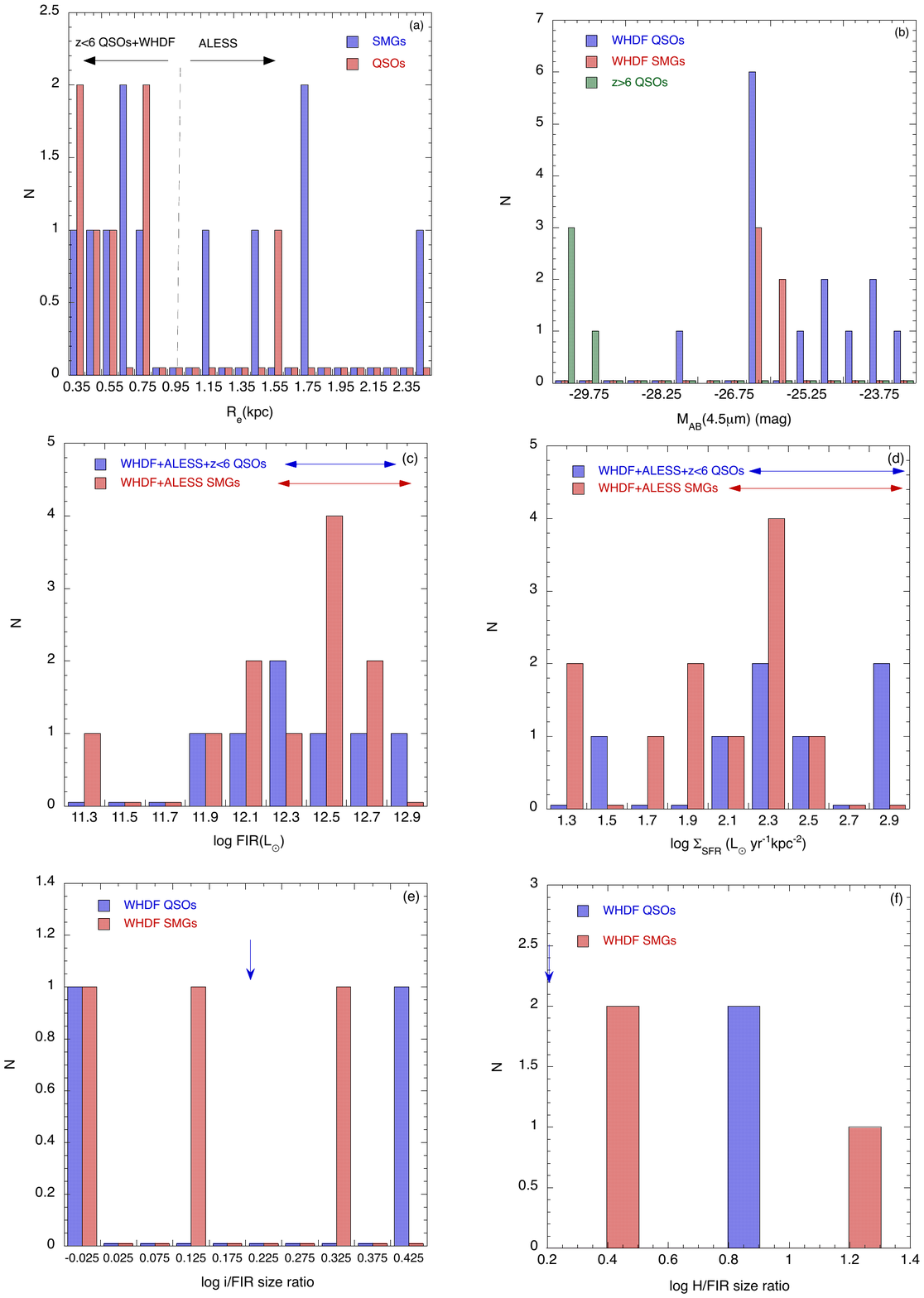}\vspace{-1.5cm}
	\caption[QSO SMG comparison]{{(a) Sub-mm effective
	radii, R$_e$, compared for QSOs and SMGs with the $z$<$6$+WHDF and
	ALESS results clearly separated on either side of the vertical
	dashed line, (b) Absolute magnitudes,
	$M_{\textup{AB}}(4.5\mu$m), compared for 14 WHDF X-ray QSOs
	(including LAB-05,-11 in $M_{\textup{AB}}(4.5\mu \rm{m})=-26.25$ mag bin), 5
	WHDF SMGs (Note that LAB-02 is undetected at 4.5$\mu$m.) and 4
	$z>6$ quasars,} (c) FIR luminosities  compared for 2 WHDF (LAB-05,
	LAB-11) + 1 ALESS (ALESS 17.1) + 4 $z>6$ quasars and 6 WHDF + 5
	ALESS SMGs{; mean $\pm1\sigma=2\sigma$ errors are shown},
	(d) SFR integrated surface densities compared for quasars and SMGs
	as in (c), (e) $i$-band/FIR continuum size ratios compared for WHDF
	quasars (LAB-05, LAB-011) and 3 SMGs (see Table
	\ref{tab:lab_size_ratios}). The arrow denotes the average
	$[\textup{C}_{\textup{II}}]$/FIR continuum size ratio measured by
	\citet{Venemans2020} for a sample of 27 $z\approx6$ QSOs, (f) Same
	as (e) for $H$-band-FIR size ratio.
	}
	\label{fig:comp6}
\end{figure*}

\section{Discussion: SMG and quasar FIR properties compared }
\label{sec:SMG_discussion}

In this work, we have presented high resolution ($0.''1$), band 7, ALMA
observations of eight $z\approx2$ sub-mm sources originally detected by
APEX LABOCA in the WHDF. {Seven of these form a flux density} limited
sample in a central $7'\times7'$ area of the WHDF. Two of these seven
WHDF sources had been previously identified as X-ray absorbed quasars at
$z=1.32$ and $z=2.12$. In addition, we presented our ALMA band 6
observations of four, $z>6$, quasars, initially detected in the VST ATLAS 
survey. We detected significant continuum dust emission in all four sources.

In \citetalias{Shanks_SMGs}, we looked for optical/NIR/MIR counterparts
for the eight WHDF SMGs detected by LABOCA and ALMA. We found that where
a counterpart was detected, most SMG SEDs were as well fitted by an
obscured AGN SED as a star-forming SED. Photometric redshifts and dust
absorptions were obtained. There were also several faint X-ray
detections, bright enough to cause a further two sources to be classed
as quasars making a minimum of 4/7 in {the flux density
limited WHDF sample}. We used the original two WHDF quasars plus the
four $z>6$ quasars as our comparison sample for the six WHDF SMGs.

All the WHDF sources were detected as resolved at $0.''1$ resolution.
Three out of four $z>6$ quasars were also more marginally resolved at
$0.''3$  angular resolution ($\approx2$kpc spatial) at $z\approx6$ in
the dust continuum emission. We find that  all these sources are small
in sub-mm extent {i.e $\approx1-2$ kpc with little
difference between the six WHDF unidentified SMGs,  the two WHDF quasars
or the four $z>6$ quasars.  They are also small relative to their host
galaxies by a factor of $\approx3$. In all cases, we find no features
that can distinguish the six unidentified WHDF SMGs as being
star-formation rather than AGN heated.}

Our comparison of the ALMA FIR continuum extent of the WHDF SMGs with
their resolved counterparts in $0.''1$ resolution HST $i$-band imaging
and $0.''9$ Calar Alto $H$-band imaging reveals that the ALMA FIR FWHM of
these objects {are generally only marginally smaller in the $i$-band by a
factor of $1.6\pm0.36$ whereas they are consistently smaller than their  $H$-band
counterparts by a factor of $7.0\pm2.4$, or $4.8\pm0.36$ if LAB-04
is excluded on the grounds of its large error.}

In this work, our primary goal was to identify the dominant fuelling
mechanism behind the observed sub-mm emission of the SMGs. To this end,
we used our ALMA observations of these sources to perform measurements
of their sizes and star formation rates based on their continuum flux densities.
We then used these measurements to calculate the star formation rate
surface densities, $\sum_{\textup{SFR}}$, of the sub-mm emitting regions
of these SMGs. We found none of these eight sources exceeded the
Eddington limit as shown as a function of R$_e$ and beam size in Fig.
\ref{fig:Hodge}. In the case of our $z>6$ SMGs we found their integrated
and peak $\sum_{\textup{SFR}}$ values to bracket the SFR densities of
the six known comparison quasars. Thus again there was little to
distinguish SMGs from known quasars in our comparison sample.  

{In passing,  we note that for five out of their sample of seven lensed, $1.5<z<3$, quasars, \cite{Stacey2021} measured small
sub-mm sizes  and correspondingly high star-formation rate densities,
similar to those found here. Their remaining two quasars  had $2-3\times$
larger extent and thus lay well below the Eddington limit. These authors
concluded that quasar sub-mm extents and implied star-formation rate densities are mostly similar to those for SMGs, in agreement with what we find here.}

We also compared our results with those of the sample of 6 ALESS
galaxies of \cite{Hodge2019}. We found one ALESS SMG that exceeded the
SFR Eddington limit so this sample now consisted of one probable quasar,
one X-ray quasar and four unidentified SMGs. In terms of SFR density the
5 other ALESS SMGs spanned similar ranges to the WHDF SMGs and their quasar
comparison sample. The major difference between the ALESS and WHDF
samples was that the ALMA observations of the ALESS SMGs studied by
\cite{Hodge2019} reveal clear signs of galactic sub-structure in these
sources, whereas we observed no signs of low-surface brightness
sub-structure in the ALMA observations of the WHDF SMGs. Within the ALESS
sample, there also seemed little to distinguish the 2 quasars from the 4
others - all showed evidence of galaxy sub-structure. However, they all
have more low surface brightness structure and larger R$_e$ than the
WHDF SMGs. Since \cite{Hodge2019} note that the only criterion used to
select these sources for high resolution ALMA observations was their
high luminosity, we conclude that this may drive this morphological
difference, with perhaps a further contribution from the ALESS
$\approx2\times$ longer  ALMA Band 7 exposures.

We note that despite the larger size of the ALESS sub-structure relative
to WHDF, their overall size is still small compared to the host galaxy
sizes we are measuring. This means there is no problem in suggesting
that these ALESS SMGs are AGN heated from the nucleus rather than from
{\it in situ} star formation.

Both samples showed little difference between the AGN and SMG
sub-samples in FIR luminosity and extent, relative size, dust masses,
SFR surface densities and NIR + MIR luminosities. Some of these
similarities are summarised in Fig. \ref{fig:comp6}. {Here,
Fig. \ref{fig:comp6}(a) compares the distribution of sub-mm effective
radii, R$_e$, of the SMGs and QSOs from the $z<6$ QSO+WHDF and ALESS
surveys. We first see that the R$_e$ from our surveys clearly separate
at R$_e\approx1$kpc from those from ALESS, with the ALESS radii being
systematically larger. We further note that within each of these
sub-samples on either side of the vertical dashed line the SMG and QSO
radii distributions appear very similar, with survey-survey systematic
differences much larger. Next,  Fig. \ref{fig:comp6} (b) now also
includes   12 other WHDF quasars in a complete X-ray sample from Table 2
of \citet{Bielby2012}. So Fig. \ref{fig:comp6} (b) shows the
similarities in the distribution of the absolute magnitudes of the WHDF
quasars and SMGs in the MIR [4.5] $\micro$m band (see Table A2 of
\citetalias{Shanks_SMGs}). Here and throughout we have assumed
$f_\lambda\propto\lambda^{-1}$ for $\lambda<4.5\micro$m as is approximately
the case for QSOs with absorption in the $0<A_V<2.5$ mag range
estimated here. This can be verified by inspecting Figs. 5 (a,b) of
\citetalias{Shanks_SMGs}. Assuming this spectral slope then implies that
the QSO  K-correction is zero for the [4.5] $\micro$m band at all redshift
and $A_V$ combinations considered here. The four $z>6$ quasars clearly
are significantly brighter due to the dominant nuclear contribution from
these extremely rare objects. The detection of the dust continuum in
these high redshift quasars does, however, show that dust can still
exist in quasar host galaxies under these conditions, close to a
hyper-massive black hole.}

{Figs. \ref{fig:comp6} (c), (d) also confirm the similarities of the FIR
luminosities and SFR densities of the WHDF+ALESS quasars and SMGs (see Table
\ref{tab:SFR}).  Even the $z>6$ quasars are indistinguishable from the 
other quasars and SMGs in both these  properties.}

Figs. \ref{fig:comp6} (e), (f) show that the ratio of sizes of the FIR
continua to the $i$-band and $H$-band host galaxy sizes are distributed
similarly for the WHDF quasars and SMGs with the $H$-band ratios being
larger (see also Table \ref{tab:lab_size_ratios}). The arrows show the
average of the ratio of FIR continuum:$[\textup{C}_{\textup{II}}]$
extents as listed in Table 3 of \cite{Venemans2020} for the 27
$z\approx6$ quasars of \citet{Decarli2018}.  These show that the
$[\textup{C}_{\textup{II}}]$ extents are generally $\approx2\times$
larger than the FIR dust continuum extents. They are also  more in
agreement with the $i$-band/FIR extent ratios than the larger
$H$-band/FIR ratios. {We note that several studies, e.g.
\cite{Simpson2015,Elbaz2018,Gullberg2019,Franco2020,Gomez2021,
Puglisi2021} argue that the small FIR size compared to the optical size
is indicative of  galaxies building their bulge. Although this scenario
cannot be ruled out, it must be said that an $\approx1$kpc radius for an
SMG powered by an AGN was predicted by many authors before the advent of ALMA
So, for example, \cite{Granato1994,Andreani1999,Kuraszkiewicz2003,Hill2011} and also
\cite{Siebenmorgen2015} predicted that any dust surrounding the central
nucleus in high-z AGN must have outer radii of $\approx1$kpc. On the
basis of these {\it a priori} predictions and on the basis of the similarity
of the distributions shown in Figs. 6 between SMGs and known quasars, we
therefore suggest that AGN are as likely to power SMGs as star-formation.

\subsection{Do MIR colours and luminosities distinguish QSOs and SMGs? }
\label{sec:MIR}}

\begin{figure*}
    \centering
	\includegraphics[width=2\columnwidth]{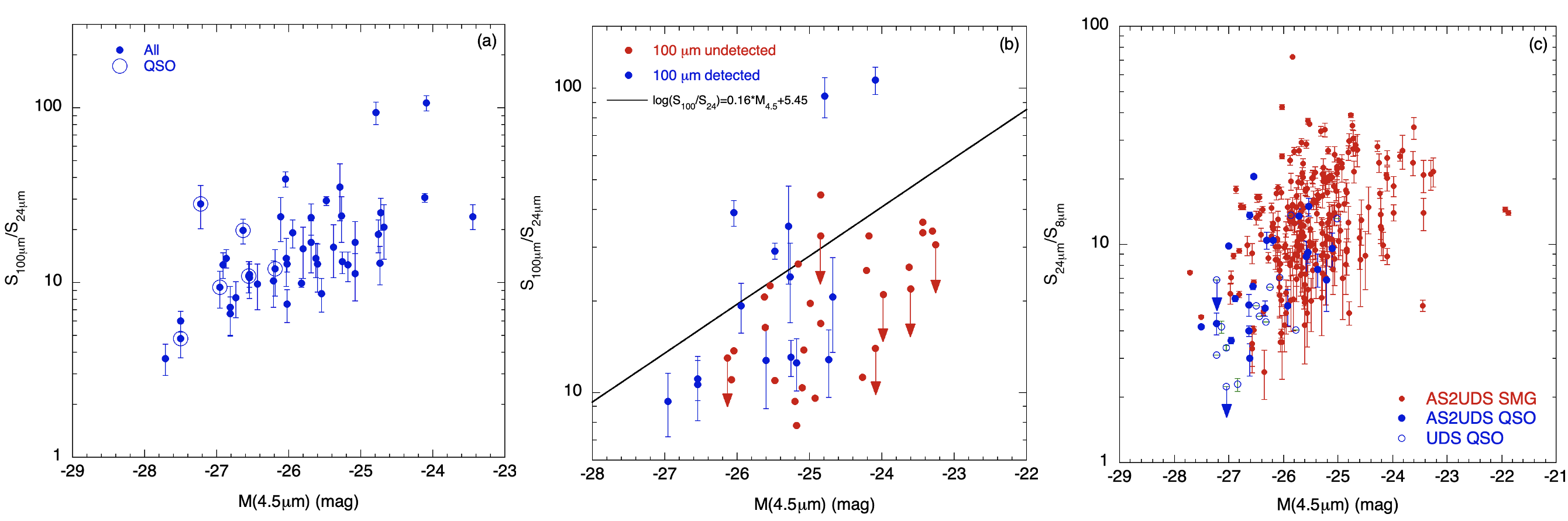}
	\caption[4.5 micron absolute magnitude versus S24/S8]{(a) $S_{100\micro
	m}/S_{24\rm \micro m}$ flux density ratios plotted against $4.5\micro$m
	absolute AB magnitudes for a sample of 48/707 SMGs that have
	detections in all 3 bands  from the AS2UDS survey of
	\protect\cite{Dudzeviciute2020}. A correlation is seen as predicted
	in \citetalias{Shanks_SMGs} to explain why bright, sub-mm loud, SDSS
	QSOs have lower $S_{70\rm \micro m}/S_{24\rm \micro m}$ ratios than other SMGs
	\citep{Hatz10}, although selection effects may  dominate this
	heavily cut sample. Also shown are 6 candidate Chandra X-ray QSOs
	\protect\citep{Kocevski2018} that are also AS2UDS detected
	(circled). These also tend to have brighter $4.5\micro$m luminosities.
	(b) 43 AS2UDS SMGs with  $1<z<1.7$ and with 24$\micro$m detections
	split  into 15 with 100$\micro$m detections (blue) and 28 with only
	100$\micro$m upper limits (red). The line is the maximum likelihood
	best fit. (c) $S_{24\rm \micro m}/S_{8\rm \micro m}$ ratios are also seen to
	correlate with $4.5\micro$m absolute  magnitudes (red filled circles)
	for 267 AS2UDS SMGs with detections in these 3 bands. 22 AS2UDS
	detected candidate  X-ray QSOs (blue filled circles) from
	\protect\cite{Kocevski2018}  and 14 K-selected UDS QSOs
	\protect\citep{Smail2008} (blue open circles) generally show
	brighter $4.5\micro$m luminosities and lower $S_{24\rm \micro m}/S_{8\rm \micro m}$
	ratios.
	}
	\label{fig:hatzim}
\end{figure*}

In \citetalias{Shanks_SMGs}, we noted that any model that implied that
bright SMGs were AGN powered still had to explain the result in Fig. 1
of \cite{Hatz10} that sub-mm-loud, broad emission line, SDSS quasars
show different Spitzer MIPS \citep{Rieke2004} $S_{70\micro m}/S_{24\rm \micro m}$
colours compared to fainter SMGs while showing similar Herschel SPIRE
\citep{Pilbratt2010,Griffin2010} $S_{350\rm \micro m}/S_{250\rm \micro m}$ colours at
longer wavelengths. In \citetalias{Shanks_SMGs}, we then predicted that
SMGs generally  should show a dependence of $S_{70\rm \micro m}/S_{24\rm \micro m}$
colours on MIR luminosity\footnote{\citetalias{Shanks_SMGs} suggested
that WHDF SMGs appearing fainter in MIR flux than  sub-mm quiet QSOs
(see their Fig. 3) was evidence for such a luminosity dependence.
However, Fig. \ref{fig:comp6}(b) shows no such difference when absolute
magnitude is considered, so no support for this hypothesis can be drawn
from Fig. 3 of \citetalias{Shanks_SMGs}. In any case, the
argument of \cite{Hatz10} applied to sub-mm-loud QSOs.}. {Unfortunately, as
noted in Sect. 2.4 of \citetalias{Shanks_SMGs}, the WHDF has no imaging
coverage between 4.5-100$\micro$m so to investigate this issue further, we
exploit the more extensive MIR coverage in the AS2UDS survey of
\cite{Dudzeviciute2020}. In Fig.~\ref{fig:hatzim}(a) we therefore plot
$S_{100\rm  \micro m}/S_{24\rm  \micro m}$ versus absolute $4.5\rm  \micro m$ magnitudes for the
48/707 AS2UDS SMGs that have detections in these 3 bands so that this
$S_{100\rm  \micro m}/S_{24\rm \micro m}$ ratio may be used as a substitute for the
$S_{70\rm  \micro m}/S_{24\rm \micro m}$ ratio used by \cite{Hatz10}. We see evidence
for a correlation  of the form required to explain the result of
\cite{Hatz10}. We further show that 6 candidate Chandra X-ray QSOs
\citep{Kocevski2018} which coincide with sub-mm sources in an AS2UDS
sub-area, also tend to show systematically brighter $4.5\rm \micro$m
luminosities and  lower  $S_{100\rm \micro m}/S_{24\rm \micro m}$ ratios.}

{We now perform a test to check if bright MIR
luminosities correlate with lower  $S_{100\rm \micro m}/S_{24\rm \micro m}$ because of
the large incompleteness (48/707) of the sample shown in
Fig.~\ref{fig:hatzim}(a). To this end, we cut the sample down to the 68
SMGs with a $24\rm \micro$m detection in the range $1<z<1.7$. At these lower
redshifts the $24\rm \micro$m sample is more complete with  43/68 having
detections. Then, of these, 15 have 100 $\rm \micro$m detections and with the
28 non-detections now also  providing more useful $S_{100\rm \micro m}/S_{24\rm \micro
m}$ upper limits, the effective detection completeness increases from
7\% to 63\%. We plot $S_{100\rm \micro m}/S_{24\rm \micro m}$ versus $M_{4.5\rm \micro m}$ in
Fig.~\ref{fig:hatzim}(b). A maximum likelihood fit of the points shown
give a linear fit of $\textup{log}_{10}(S_{100\rm \micro m}/S_{24\micro
m})=0.16\pm0.03\times M_{4.5}+5.45\pm0.3$ with the correlation
significant at $\approx3\sigma$, supporting the reality of the apparent
correlation seen in Fig.~\ref{fig:hatzim}(a). Moreover, in
Fig.~\ref{fig:hatzim}(c) we similarly show $S_{24\rm \micro m}/S_{8\rm \micro m}$
versus $M_{4.5\rm \micro m}$ for 267/707 AS2UDS sources detected in all these 3
bands. Here the $24\micro$m and $8\micro$m data come from the $\approx1$
deg$^2$ {\it SpUDS} survey (PI J. Dunlop). We find that this 38\%
completeness is at least enough to confirm the trend that  brighter
$4.5\rm \micro m$ SMGs have lower  ratios in these bands,  similar to the above
$S_{100\rm \micro m}/S_{24\rm \micro m}$ results. We also plot 22 candidate QSOs from
the X-UDS Chandra X-ray 0.33 deg$^2$ survey of \cite{Kocevski2018} that
are also listed in AS2UDS and have $[3.6\rm \micro m]-[4.5\rm \micro m]>0.5$ mag
(Vega), an established MIR criterion for QSO selection
\citep{Stern2012}. AS2UDS photometric redshifts are adopted for these.
Also shown are 14 QSOs selected by K-excess in the UDS field by
\cite{Smail2008}, although only 3 are detected at $24 \rm \micro m$ and the
other 11 represent $S_{24\rm \micro m}/S_{8\rm \micro m}$ upper limits in
Fig.~\ref{fig:hatzim}(c). 4 of these UDS QSOs overlap with the 22
candidate AS2UDS X-ray QSOs. We again see an apparent correlation
between $4.5\micro$m luminosity and now the $S_{24\rm \micro m}/S_{8\rm \micro m}$ ratio
with both QSO samples populating the bright end of the distribution. We
conclude first that the result in Fig.~\ref{fig:hatzim}(c) is in line
with QSOs having lower $S_{70\rm \micro m}/S_{24\rm \micro m}$ ratios than fainter
SMGs as found by  \cite{Hatz10}. Second, we conclude that the results in
Figs.~\ref{fig:hatzim}(a,b,c) all appear to suggest that sub-mm sources
with  brighter $4.5\micro$m luminosities have  lower $S_{100\rm \micro m}/S_{24\micro
\textup{m}}$ and $S_{24\rm \micro m}/S_{8\rm \micro m}$ ratios. This result is naturally 
interpreted if more MIR luminous SMGs and sub-mm-loud QSOs 
have both hot and cold dust components, compared to less luminous sources
that only show a cold dust component.
\smallskip

\subsection{Does the presence of low surface-brightness sub-structure depend on SMG FIR luminosity?}}
\label{sec:LSB}

Finally, we also try to explain the absence in all of the WHDF sample of
low surface brightness sub-structure such as spiral arms seen in the
sample of six ALESS SMGs of \cite{Hodge2019}. Again we suggest that this
could be a luminosity effect but here with  the FIR sub-mm luminosity
driving the different morphology seen in the higher luminosity SMG
sample. However, some further contribution may arise from lower S/N in
the WHDF sub-mm sample compared to ALESS. This latter possibility can be
further checked by significantly increasing the ALMA exposure time on
the WHDF SMG sample. Either way, the ALESS result can still be explained
by AGN nuclear heating, rather than {\it in situ} star-formation,
heating the spiral arms since their extent is still small compared to
the optical/NIR/[C{\sc ii}] extent of SMG host galaxies.
\smallskip

\section{Conclusions}
\label{sec:SMG_conclusions}

We have observed eight WHDF sub-mm sources including two known absorbed X-ray quasars 
at $0.''1$ resolution in ALMA and  {four} VST ATLAS $z>6$ quasars at $0.''3$
resolution and resolved all but one $z>6$ quasar in the dust continuum.
{Our conclusions are as follows.}

\begin{enumerate}

\item  {As measured in the  MIR (e.g. [4.5]$\micro$m),} the intrinsic
luminosities of the  WHDF SMGs and quasars in the {sub-mm flux density limited} 
sample are similar. They are also in the same range as mostly unabsorbed
X-ray quasars in the complete WHDF X-ray sample that partly overlaps the sub-mm sample. 
\smallskip

\item  All the sub-mm sizes of the WHDF SMGs are {compact 
($\approx1-2$kpc) with no difference in physical} size 
compared to the two WHDF sub-mm-loud quasars or the four $z>6$ quasars.
\smallskip

\item All the sub-mm sizes of the WHDF SMGs are {compact
relative to the size of the  host galaxy} when detected in the optical
or NIR. Again there is no difference in sub-mm - host relative extents
between the six unidentified SMGs and the six quasars.
\smallskip

\item There is also little  difference in either FIR luminosity or SFR
surface density between the six WHDF SMGs and the six quasars.  
{All lie below the SFR density `Eddington limit' except for one $z>6$ 
quasar (and one ALESS SMG).}
\smallskip

\item There {\it is} a difference between {our ALMA
observations of the eight WHDF SMGs } and the six ALESS SMGs observed by
\cite{Hodge2019} in that the WHDF sources show no evidence of the low
surface brightness spiral arms found in the ALESS sample. Deeper ALMA
observations of the WHDF SMGs will show if this is due to the higher S/N
of the ALESS sample or whether it is due to the ALESS SMG's higher
sub-mm luminosity. Either way, the scale of the spiral arms remains  so
small that they are still consistent with AGN heating from the nucleus.

\smallskip



\item {We find  evidence in the  AS2UDS survey
\citep{Dudzeviciute2020} that sub-mm loud QSOs show lower $S_{100\micro
m}/S_{24\rm \micro m}$ and $S_{24\rm \micro m}/S_{8\rm \micro m}$ flux density ratios than
other SMGs, similar to the result of \cite{Hatz10}. We also find
preliminary evidence of a correlation between these flux density ratios 
and MIR $4.5\rm \micro m$ luminosity. If this correlation is confirmed then it
will suggest that more MIR luminous SMGs and sub-mm loud QSOs include
hot as well as cold dust components. It would also remove the objection,
based on the \cite{Hatz10} result, to the idea that most SMGs with
$S_{870\micro{\rm m}}\ga 3$ mJy are AGN powered.}

\end{enumerate}

{To these can be added the broad conclusion from
\citetalias{Shanks_SMGs} that the  unidentified WHDF SMGs with
optical/NIR/MIR counterparts can be  as well fitted by AGN SEDs as by
star-forming galaxy SEDs. It should also be noted that the compact SMG
sizes found by ALMA were a clear {\it a priori} prediction unique to
the AGN powered SMG model. Indeed, the observed lack of dust heating at radii
larger than 1-2kpc in galaxies that extend to $\ga8$ kpc could be taken
as a strong signature for nuclear as opposed to {\it in situ}  heating
by local star-formation.} Our overall conclusion is therefore  that
there is no fundamental argument against AGN heating SMG dust rather
than star formation. Indeed, the similarities between the AGN and SMG
populations positively suggest that AGN heating may dominate at least in
the brightest SMGs, allowing these dust-obscured quasars to help explain
both the hard X-ray and FIR cosmic backgrounds.

\section*{Acknowledgements}
We thank I.R. Smail for specific suggestions to improve Section
\ref{sec:MIR}. We also thank  an anonymous referee for extensive
suggestions that improved the quality of this paper throughout. BA acknowledges support from the Australian Research Council’s Discovery Projects scheme (DP200101068). This
paper makes use of the following ALMA data: ADS/JAO.ALMA-2016.1.01523.S
and ADS/JAO.ALMA-2016.1.01510.S. ALMA is a partnership of ESO
(representing its member states), NSF (USA) and NINS (Japan), together
with NRC (Canada), MOST and ASIAA (Taiwan), and KASI (Republic of
Korea), in cooperation with the Republic of Chile. The Joint ALMA
Observatory is operated by ESO, AUI/NRAO and NAOJ.

\section*{Data Availability} 
The catalogue data underlying this article are available in the article and  Appendix A of \citetalias{Shanks_SMGs}.
The imaging data underlying this article are publicly available in the ALMA archive.




\bibliographystyle{mnras}
\bibliography{Bibliography} 








\bsp	
\label{lastpage}
\end{document}